\documentclass[twocolumn]{aastex63}

\usepackage{courier}

\DeclareRobustCommand{\ion}[2]{%
\relax\ifmmode
\ifx\testbx\f@series
{\mathbf{#1\,\mathsc{#2}}}\else
{\mathrm{#1\,\mathsc{#2}}}\fi
\else\textup{#1\,{\mdseries\textsc{#2}}}%
\fi}

\usepackage{scalerel}
\newcommand\HI{H\protect\scaleto{$I$}{1.2ex}}

\usepackage{graphicx}
\usepackage{courier}
\usepackage{amsmath} 
\usepackage{gensymb}
\usepackage{booktabs}

\usepackage{longtable}
\usepackage{float}

\shorttitle{Local kpc-scale properties of stars, dust, and gas in nearby spiral galaxies}
\shortauthors{Abdurro'uf et al.}
\graphicspath{{./}{figures/}}

\begin{document}

\title{Dissecting Nearby Galaxies with \texttt{piXedfit}: I. Spatially Resolved Properties of Stars, Dust, and Gas as Revealed by Panchromatic SED Fitting}

\correspondingauthor{Abdurro'uf}
\email{abdurrouf@asiaa.sinica.edu.tw}

\author[0000-0002-5258-8761]{Abdurro'uf}
\affiliation{Institute of Astronomy and Astrophysics, Academia Sinica, \\
11F of AS/NTU Astronomy-Mathematics Building, No.1, Sec. 4, Roosevelt Road, Taipei 10617, Taiwan, R.O.C.}

\author[0000-0001-7146-4687]{Yen-Ting Lin}
\affiliation{Institute of Astronomy and Astrophysics, Academia Sinica, \\
11F of AS/NTU Astronomy-Mathematics Building, No.1, Sec. 4, Roosevelt Road, Taipei 10617, Taiwan, R.O.C.}

\author[0000-0002-4189-8297]{Hiroyuki Hirashita}
\affiliation{Institute of Astronomy and Astrophysics, Academia Sinica, \\
11F of AS/NTU Astronomy-Mathematics Building, No.1, Sec. 4, Roosevelt Road, Taipei 10617, Taiwan, R.O.C.}

\author[0000-0002-8512-1404]{Takahiro Morishita}
\affiliation{Space Telescope Science Institute, 3700 San Martin Drive, Baltimore, MD 21218, USA}

\author[0000-0002-8224-4505]{Sandro Tacchella}
\affiliation{Department of Physics, Ulsan National Institute of Science and Technology (UNIST), Ulsan 44919, Republic of Korea}

\author[0000-0002-2651-1701]{Masayuki Akiyama}
\affiliation{Astronomical Institute, Tohoku University, Aramaki, Aoba, Sendai 980-8578, Japan}

\author[0000-0001-8416-7673]{Tsutomu T.\ Takeuchi}
\affiliation{Division of Particle and Astrophysical Science, Nagoya
University, Furo-cho, Chikusa-ku, Nagoya 464--8602, Japan}
\affiliation{The Research Center for Statistical Machine Learning, the
Institute of Statistical Mathematics, \\ 10-3 Midori-cho, Tachikawa, Tokyo
190---8562, Japan}

\author[0000-0002-9665-0440]{Po-Feng Wu}
\affiliation{Institute of Astronomy and Astrophysics, Academia Sinica, \\
11F of AS/NTU Astronomy-Mathematics Building, No.1, Sec. 4, Roosevelt Road, Taipei 10617, Taiwan, R.O.C.} 
\affiliation{National Astronomical Observatory of Japan, 2-21-1 Osawa, Mitaka, Tokyo 181-8588, Japan}



\begin{abstract}

We study spatially resolved properties (on spatial scales of $\sim 1-2$ kpc out to at least $3$ effective radii) of the stars, dust, and gas in ten nearby spiral galaxies. The properties of the stellar population and dust are derived by fitting the spatially resolved spectral energy distribution (SED) with more than 20 photometric bands ranging from far-ultraviolet to far-infrared. Our newly developed software \verb|piXedfit| performs point spread function matching of images, pixel binning, and models the stellar light, dust attenuation, dust emission, and emission from a dusty torus heated by an active galactic nucleus simultaneously through the energy balance approach. With this self-consistent analysis, we present the spatially resolved version of the IRX--$\beta$ relation, finding that it is consistent with the relationship from the integrated photometry. We show that the old stellar populations contribute to the dust heating, which causes an overestimation of star formation rate (SFR) derived from the total ultraviolet and infrared luminosities on kpc scales. With archival high-resolution maps of atomic and molecular gas, we study the radial variation of the properties of the stellar populations (including stellar mass, age, metallicity, and SFR), dust (including dust mass, dust temperature, and abundance of polycyclic aromatic hydrocarbon), gas, as well as dust-to-stellar mass and dust-to-gas mass ratios. We observe a depletion of molecular gas mass fraction in the central region of the majority of the galaxies, suggesting that the lack of available fuel is an important factor in suppressing the specific SFR at the center. 

\end{abstract}



\section{Introduction} \label{sec:intro}
A galaxy is a complex system whose baryonic component is composed of stars, dust, gas, and metals. These four key components are correlated with each other in a complex manner. Stars are formed from the cold molecular gas. During the lives of stars, metals are produced in the process of nucleosynthesis. Metals are ejected from stars to the interstellar medium (ISM) through the stellar wind and supernovae (SNe). Metals are also the main ingredient for dust formation. The dust grains form through the condensation of metals in the stellar ejecta \citep[e.g., ][]{1998Dwek, 2008Zhukovska} that include winds of low mass stars in the asymptotic giant branch phase \citep{2009Valiante,2006Ferrarotti} and the Type II SNe at the end of the life of massive stars \citep{2001Todini,2003Nozawa}. During their evolution, dust grains in the dense molecular clouds can grow through accretion and coagulation \citep[][]{1993Ossenkopf, 2009Ormel, 2011Hirashita}. Dust can be destroyed through SNe shocks \citep[e.g., ][]{2008Zhukovska, 2015Slavin}, thermal evaporation, cosmic rays, and dust inclusion into star formation \citep[e.g., ][]{1998Dwek, 1999Hirashita}. Aside from modifying the spectral energy distribution (SED) of galaxies, dust also plays important roles in the evolution of galaxies by catalyzing the formation of molecular hydrogen \citep{1979Hollenbach} and shielding gas from the interstellar radiation field (ISRF). These effects help the gas cooling and produce a favorable condition for star formation \citep{1999Hollenbach, 2009Krumholz, 2011Krumholz, 2011Yamasawa, 2012Glover}.

The four baryonic components mentioned above contribute differently to the integrated SED of a galaxy. The stellar component emits light mainly in the far-ultraviolet (FUV) to near-infrared (NIR) regime. Dust absorbs the FUV--optical light and re-emits the light as thermal radiation in the mid-infrared (MIR) to far-infrared (FIR) regime. Metals in the stellar atmosphere contribute to the reddening of the SED in the optical--NIR. Hot gas ($>10^{6}$ K) emits strongly in the X-ray and ultraviolet (UV), while neutral and molecular gas are observed through emission lines in the submillimeter and radio. Multi-wavelength observations are thus required to study these components and compare their relative significance in the evolution of galaxies. Standard stellar population synthesis techniques then provide a way for extracting (i.e.,~decomposing) the properties of those components \citep[e.g., ][]{2011Walcher, 2013Conroy}.          

Nearby galaxies are a very useful laboratory for studying the interplay among the aforementioned baryonic components because their proximity enables us to obtain detailed, spatially resolved information and to conduct comprehensive analyses of these components, thanks to the abundance of imaging and spectroscopic surveys ranging from FUV to radio. Many studies have investigated the spatially resolved properties of the stellar, dust, and gas components in nearby galaxies. They used various methods to derive the properties of the stellar population and dust, while the surface densities of gas (both neutral and molecular components) were derived using the \HI{} and CO data. These studies can roughly be classified into four categories. 

In the first category, which accounts for the majority of the studies, the imaging data from MIR to FIR are used and the spatially resolved FIR dust emission SEDs are fitted with the modified black body models to derive the spatially resolved properties of dust (e.g.,~mass and temperature), while the star formation rate (SFR) is derived using MIR photometry alone \citep[e.g.,][]{2010Bendo, 2012Bendo, 2010Braine, 2011Boquien, 2012Foyle, 2012Groves, 2012Smith, 2012Xilouris, 2013Meixner, 2014Gordon, 2014Roman-Duval, 2017Chastenet, 2018Chiang, 2019Utomo}. The second category of the studies used UV and MIR imaging data and derived the SFR using a simple prescription that combines data at these two wavelengths, while the \HI{} and CO data are used for deriving the surface densities of gas \citep[e.g.,][]{2008Bigiel}. However, these studies did not analyze dust properties. Studies in the third category have used the multiwavelength imaging data sets covering UV to FIR. However, the properties of the stellar population and dust are not derived using a single method \citep[e.g., the stellar population synthesis method;][]{2009Munoz-Mateos_b, 2017Casasola, 2017Eufrasio, 2020Enia, 2020Morselli, 2021Wei}. \citet{2017Casasola} used FUV--FIR data along with the \HI{} and CO data to analyze the radial distributions of dust, stars, gas, and SFR. In their analysis, dust mass is obtained by modeling the FIR dust emission SED, while the SFR and $M_{*}$ are derived using simple prescriptions as mentioned above.   

Finally, studies in the fourth category have used a single method, namely the spatially resolved SED modeling that adopts the energy balance approach (i.e.,~the energy that is absorbed by dust in the UV to NIR is equal to the energy that is re-radiated in the infrared (IR); \citealt{2008daCunha}, \citealt{2009Noll}, \citealt{2017Leja}) to derive the properties of stellar population and dust \citep[e.g.,][]{2012Boquien, 2014DeLooze, 2017Viaene, 2019Williams, 2020Nersesian_a}. The energy balance approach has been shown to work well on kpc scales by previous studies. \citet{2014DeLooze} and \citet{2019Williams} performed a high-resolution 3D radiative transfer modeling to NGC 5194 (M51a) and NGC 598 (M33), respectively, to reconstruct the spatially resolved multiwavelength fluxes of these galaxies. They demonstrated that the energy balance assumption is valid on scales of $\gtrsim 1$ kpc. This is in agreement with the study by \citet{2018Smith} who conducted a pixel-by-pixel panchromatic SED fitting experiment on a simulated disk galaxy. They found that it is increasingly more difficult to recover the true spatially resolved parameters of the galaxy when the pixel scale is $\lesssim 1$ kpc owing to the inaccuracy of the energy balance assumption in this regime.

Although there have been many studies on nearby galaxies, only a few of them have actually studied the spatially resolved properties of the stellar, dust, and gas components simultaneously. Therefore, we often need to piece together results obtained from various studies in order to understand the interplay among these baryonic components in the galaxies. Some of the key results revealed by the previous works are summarized below. By analyzing the spatially resolved SFR and gas mass, \citet{2008Bigiel} found a local kpc analog of the Kennicutt-Schmidt (KS) relation \citep{1959Schmidt, 1998Kennicutt}, which is a relation between the SFR surface density ($\Sigma_{\rm SFR}$) and the molecular gas mass surface density ($\Sigma_{\rm{H}_2}$), with a slope of $\sim 1$, indicating a relatively constant star formation efficiency (SFE). These authors estimated the average molecular gas depletion time to be $\sim 2$ Gyrs. \citet{2009Munoz-Mateos_b} observed a decreasing dust-to-gas mass ratio with radius in the majority of their sample of nearby galaxies. \citet{2012MentuchCooper} found a broadly constant dust-to-stellar mass ratio in NGC 5194 with an average of $\log(M_{\rm dust}/M_{*})=-2.5\pm 0.2$, suggesting an evolutionary coupling between the stellar and dust components. They also found that the dust in the central region of NGC 5195 is heated by a strong ISRF up to a temperature of $\sim 30$ K. They proposed some heating mechanisms from old stellar populations and an active galactic nucleus (AGN). Other studies have also found indications of the dust heating by old stellar populations in NGC 5195 \citep[e.g.,~][]{2017Eufrasio, 2020Nersesian_b}.

Recent studies of the spatially resolved dust properties from the DustPedia project \citep{2017Davies} that combined FUV--FIR imaging data and applied a 3D radiative transfer modeling have found a significant contribution ($\sim 20-90\%$) of the old stellar population to the dust heating  \citep[e.g.,][]{2014DeLooze, 2017Viaene, 2020Nersesian_a, 2020Nersesian_b}. These results suggest that the SFR estimation through a conversion from the infrared (IR) luminosity has to be used with caution. A correction for the dust heating by the old stellar population needs to be applied in order to obtain a robust SFR estimate. In line with these results, previous studies have found that the SFR calculated using simple prescriptions that combines UV and IR luminosities can possibly overestimate the true SFR owing to the uncorrected dust heating by the old stellar population in the simple prescriptions \citep[e.g.,][]{2003Hirashita, 2019Leja}. The overestimation of the SFR becomes severe for passive galaxies where old stars are the dominant population.

In this paper, which is the first in a series, we study the spatially resolved distributions and properties of the stellar, dust, and gas components in ten nearby spiral galaxies using panchromatic imaging data that cover FUV--FIR wavelength range. The sample galaxies cover a relatively wide range of specific SFR, which allows us to study the interplay among these baryonic components in various star formation phases. In accompanying papers, we will study scaling relations in local kpc scales and the resolved distributions of the dust-to-gas mass and dust-to-metals ratios. In this work, we employ our newly-developed software \texttt{piXedfit} \citep{2021Abdurrouf} for conducting the spatially resolved panchromatic SED fitting to self-consistently derive the properties of stellar population and dust. 

Our analysis have at least two new contributions: (1) the usage of a single method (i.e.,~the spatially resolved SED fitting) for deriving the properties of stellar population and dust. By using the SED fitting method that adopts the energy-balance approach, we expect the analysis to be more self-consistent than the ones that use two different methods for the stellar population and dust (e.g.,~dust properties are derived by fitting the FIR SED with the modified black body models, while the SFR and stellar mass are obtained using simple prescriptions). (2) The rich data sets and the sufficient number of galaxies analyzed in this study enable a comprehensive study of the major baryonic components in galaxies. 

This paper is organized as follows. We describe the data and sample selection in Section~\ref{sec:data}. The methodology is presented in Section~\ref{sec:method}. In Section~\ref{sec:results}, we show our results that include maps of various components, comparison between the SFR derived with \verb|piXedfit| and that estimated using the total UV and IR luminosities, resolved IRX--$\beta$ relations, and the radial profiles of the stellar, dust, and gas components. In Section~\ref{sec:discussion}, we further investigate the relative distributions of the stellar, dust, and gas components. At the end of this section, we discuss the possible effects of the CO-to-$\text{H}_{2}$ conversion factor on the main results presented in this paper. Finally, we summarize our results in Section~\ref{sec:summary}.

Throughout this paper, we assume the \citet{2003Chabrier} initial mass function (IMF), cosmological parameters of $\Omega_{m}=0.3$, $\Omega_{\Lambda}=0.7$, and $H_{0}=70\text{ km}\text{ s}^{-1}\text{ Mpc}^{-1}$, and the AB magnitude system.

\section{Data and Sample Selection} \label{sec:data}
In the analysis thoughout this paper, we use broad-band imaging data ranging from the FUV to FIR, \HI{} 21 cm line emission map, and CO intensity map. In this section, we describe each of the data sets. 

\subsection{Broad-band FUV--FIR Imaging Data} \label{sec:broad_band_images}
We use archival multiband imaging data of 24 bands collected from various surveys that make use of both the space-based and ground-based telescopes, including the Galaxy Evolution Explorer \citep[GALEX;][]{2005Martin,2007Morrissey}, the Sloan Digital Sky Survey \citep[SDSS;][]{2000York}, the Two Micron All Sky Survey \citep[2MASS;][]{2006Skrutskie}, the Wide-field Infrared Survey Explorer \citep[WISE;][]{2010Wright}, the \textit{Spitzer Space Telescope} \citep{2003Gallagher}, and the \textit{Herschel Space Observatory} \citep{2010Pilbratt}.

Imaging data in the FUV and near-ultraviolet (NUV) bands are taken from the GALEX surveys, including the Nearby Galaxies Survey \citep[NGS;][]{2003Bianchi_a,2003Bianchi_b,2004GildePaz}, Guest Investigators Survey (GII), and All-sky Imaging Survey (AIS). The imaging data in the optical $u$, $g$, $r$, $i$, and $z$ bands are taken from the SDSS DR12\footnote{\url{https://dr12.sdss.org/mosaics/}} \citep{2015Alam}. The near-infrared (NIR) imaging data in the $J$, $H$, and $K_{s}$ bands are taken from the 2MASS Large Galaxy Atlas\footnote{\url{https://irsa.ipac.caltech.edu/applications/2MASS/LGA/}} \citep{2003Jarrett}. The mid-infrared (MIR) part of the spectral energy distribution (SED) is covered by the WISE and \textit{Spitzer}. The WISE imaging data in the $W1$, $W2$, $W3$, and $W4$ bands are taken from the AllWISE data release\footnote{The NASA/IPAC Infrared Science Archive (IRSA) \url{https://irsa.ipac.caltech.edu/applications/wise/}} \citep{2013Cutri} which combines data from the WISE cryogenic and the NEOWISE \citep{2011Mainzer} post-cryogenic survey phases. In this work, we use imaging data in the four bands of IRAC (the Infrared Array Camera; \citealt{2004Fazio}; $3.6$, $4.5$, $5.8$, and $8.0$ $\mu$m) and one band of MIPS (the Multiband Imaging Photometer for Spitzer; \citealt{2004Rieke}; $24$ $\mu$m). The \textit{Spitzer} imaging data in the four IRAC bands are taken from the \textit{Spitzer} Infrared Nearby Galaxies Survey \citep[SINGS\footnote{\url{https://irsa.ipac.caltech.edu/data/SPITZER/SINGS/}};][]{2003Kennicutt} and \textit{Spitzer} Enhanced Imaging Products (SEIP), while the imaging data in MIPS $24$ $\mu$m are taken from the SINGS and Local Volume Legacy \citep[LVL;][]{2008Kennicutt, 2009Lee, 2009Dale} surveys. The imaging data in the FIR are taken from the Key Insights on Nearby Galaxies: a Far-Infrared Survey with \textit{Herschel} \citep[KINGFISH\footnote{\url{https://irsa.ipac.caltech.edu/data/Herschel/KINGFISH/}};][]{2011Kennicutt} and Very Nearby Galaxy Survey \citep[VNGS\footnote{\url{https://irsa.ipac.caltech.edu/data/Herschel/VNGS/index.html}};][]{2012Bendo} which conducted the observations using the \textit{Herschel Space Observatory}. The FIR imaging data used in this analysis consists of three PACS \citep[Photodetector Array Camera and Spectrometer;][]{2010Poglitsch} bands ($70$, $100$, and $160$ $\mu$m) and two SPIRE \citep[Spectral and Photometric Imaging Receiver;][]{2010Griffin} bands ($250$ and $350$ $\mu$m). The list of imaging data used in this work and their basic characteristics are summarized in Table~\ref{tab:filters_set}. 

Next we describe the limiting magnitudes and surface brightness of the imaging data. The $5\sigma$ limiting magnitudes in FUV (NUV) of the three main survey modes in GALEX, AIS, Medium Imaging Survey (MIS), and Deep Imaging Survey (DIS) are $19.9$ ($20.8$), $22.6$ ($22.7$), and $24.8$ ($24.4$), respectively \citep{2007Morrissey}. The NGS survey has similar limiting magnitudes to the MIS survey, while the GII program is comparable with the DIS. The SDSS imaging is 95$\%$ complete to $u=22.0$ mag, $g=22.2$ mag, $r=22.2$ mag, $i=21.3$ mag, and $z=20.5$ mag \citep{2004Abazajian}.  The point-source sensitivities at a signal-to-noise ratio (S/N) of $10$ are: $15.8$, $15.1$, and $14.3$ mag for $J$, $H$, and $K_{s}$ bands, respectively. The WISE achieves $5\sigma$ point source sensitivites better than $0.08$, $0.11$, $1$, and $6.0$ mJy in unconfused regions on the ecliptic in the $W1$, $W2$, $W3$, and $W4$ bands, respectively \citep{2010Wright}. The $3\sigma$ surface brightness limits in the IRAC and MIPS bands are $0.02$ $\text{MJy}\text{ sr}^{-1}$ ($3.6$ $\mu$m), $0.03$ $\text{MJy}\text{ sr}^{-1}$ ($4.5$ $\mu$m), $0.09$ $\text{MJy}\text{ sr}^{-1}$ ($5.8$ $\mu$m), $0.12$ $\text{MJy}\text{ sr}^{-1}$ ($8.0$ $\mu$m), and $0.2$ $\text{MJy}\text{ sr}^{-1}$ ($24$ $\mu$m) \citep{2003Kennicutt}. The $1\sigma$ sensitivity per pixel for bright sources in the \textit{Herschel} PACS and SPIRE bands are $7.1$, $7.1$, $3.1$, $1.0$, $0.6$, and $0.3$ $\text{MJy}\text{ sr}^{-1}$ at $70$, $100$, $160$, $250$, $350$, and $500$ $\mu$m, repsectively \citep{2011Kennicutt}. 

\begin{deluxetable*}{ccccccccc}
\tablenum{1}
\tablecaption{Basic Information of the Broad-band FUV--FIR Imaging Data\label{tab:filters_set}}
\tablewidth{0pt}
\tablehead{
\colhead{} & \colhead{} & \colhead{} &  \colhead{Dataset} & \colhead{Central} & \colhead{PSF\textsuperscript{a}} & \colhead{Pixel} & \multicolumn{2}{c}{Variation of spatial resolution (SR)\textsuperscript{b}} \\
\cline{8-9}
\colhead{Telescope} & \colhead{Instrument} &  \colhead{Filter} &  \colhead{or survey} & \colhead{Wavelength} & \colhead{FWHM} & \colhead{Resolution} & \colhead{SR1 ($24.88''$)} & \colhead{SR2 ($11.89''$)} \\
\colhead{} & \colhead{} & \colhead{} &  \colhead{} & \colhead{($\mu$m)} & \colhead{(arcsec)} & \colhead{(arcsec)} & \colhead{(flag)} & \colhead{(flag)}
}
\decimals
\startdata
GALEX & \nodata & FUV & NGS, GII, AIS &$0.1516$ & $4.48$ & $1.5$ & 1 & 1 \\
GALEX & \nodata & NUV & NGS, GII, AIS & $0.2267$ & $5.05$ & $1.5$ & 1 & 1 \\
SDSS & \nodata & $u$ & DR12 & $0.3551$ & $1.5$ & $0.396$ & 1 & 1 \\
SDSS & \nodata & $g$ & DR12 & $0.4686$ & $1.5$ & $0.396$ & 1 & 1 \\
SDSS & \nodata & $r$ & DR12 & $0.6166$ & $1.5$ & $0.396$ & 1 & 1 \\
SDSS & \nodata & $i$ & DR12 & $0.7480$ & $1.0$ & $0.396$ & 1 & 1 \\
SDSS & \nodata & $z$ & DR12 & $0.8932$ & $1.0$ & $0.396$ & 1 & 1 \\
2MASS & \nodata & $J$ & LGA & $1.2358$ & $3.5$ & $1.0$ & 1 & 1 \\
2MASS & \nodata & $H$ & LGA & $1.6458$ & $3.5$ & $1.0$ & 1 & 1 \\
2MASS & \nodata & $K_{s}$ & LGA & $2.1603$ & $3.5$ & $1.0$ & 1 & 1 \\
WISE & \nodata & $W1$ & AllWISE & $3.3526$ & $5.79$ & $1.375$ & 1 & 1 \\
WISE & \nodata & $W2$ & AllWISE & $4.6028$ & $6.37$ & $1.375$ & 1 & 1 \\
WISE & \nodata & $W3$ & AllWISE & $11.5608$ & $6.60$ & $1.375$ & 1 & 1 \\
WISE & \nodata & $W4$ & AllWISE & $22.0883$ & $11.89$ & $1.375$ & 1 & 1 \\
\textit{Spitzer} & IRAC & $3.6$ $\mu$m & SINGS, SEIP & $3.5569$ & $1.90$ & $1.2/0.6$\textsuperscript{c} & 1 & 1 \\
\textit{Spitzer} & IRAC & $4.5$ $\mu$m & SINGS, SEIP &  $4.5020$ & $1.81$ & $1.2/0.6$\textsuperscript{c} & 1 & 1 \\
\textit{Spitzer} & IRAC & $5.8$ $\mu$m & SINGS, SEIP &  $5.7450$ & $2.11$ & $1.2/0.6$\textsuperscript{c} & 1 & 1 \\
\textit{Spitzer} & IRAC & $8.0$ $\mu$m &  SINGS, SEIP &  $7.9158$ & $2.82$ & $1.2/0.6$\textsuperscript{c} & 1 & 1 \\
\textit{Spitzer} & MIPS & $24$ $\mu$m & SINGS, LVL &  $23.2096$ & $6.43$ & $1.5$ & 1 & 1 \\
\textit{Herschel} & PACS & $70$ $\mu$m & KINGFISH, VNGS &  $68.9247$ & $5.67$ & $1.4$ & 1 & 1 \\
\textit{Herschel} & PACS & $100$ $\mu$m & KINGFISH, VNGS &  $97.9036$ & $7.04$ & $1.7$ & 1/0\textsuperscript{d}  & 1/0\textsuperscript{d} \\
\textit{Herschel} & PACS & $160$ $\mu$m & KINGFISH, VNGS &  $153.9451$ & $11.18$ & $2.85$ & 1 & 1  \\
\textit{Herschel} & SPIRE & $250$ $\mu$m & KINGFISH, VNGS &  $247.1245$ & $18.15$ & $6.0$ & 1 & 0 \\
\textit{Herschel} & SPIRE & $350$ $\mu$m & KINGFISH, VNGS &  $346.7180$ & $24.88$ & $10.0/8.0$\textsuperscript{e} & 1 & 0 \\
\enddata
\tablecomments{
\textsuperscript{a} The point spread function (PSF) full width at half-maximum (FWHM) in each band is taken from \citet{2011aniano}. Because we use the convolution kernels from \citet{2011aniano} for the PSF matching process (to be described in Section~\ref{sec:image_process}), for consistency, we refer to the PSF FWHM values from \citet{2011aniano}. 
\textsuperscript{b} Flag indicating whether an image is used or not in the two sets of imaging data for varying the spatial resolution (SR1 and SR2; see Section~\ref{sec:image_process}). 
\textsuperscript{c} SINGS image has $1.2''\text{ pixel}^{-1}$, while SEIP image has $0.6''\text{ pixel}^{-1}$. 
\textsuperscript{d} For a galaxy that is observed by the VNGS survey (see Table~\ref{tab:surveys_GLX_IRC_MIPS}), the PACS $100$ $\mu$m band is not available. 
\textsuperscript{e} Imaging data from the KINGFISH and VNGS surveys have a spatial sampling of $10''\text{ pixel}^{-1}$ and $8.0''\text{ pixel}^{-1}$, respectively.}
\end{deluxetable*}

\subsection{\HI{} and CO Maps} \label{sec:radio_images}
We use \HI{} and CO maps taken from The \HI{} Nearby Galaxy Survey \citep[THINGS\footnote{\url{https://www2.mpia-hd.mpg.de/THINGS/Overview.html}};][]{2008Walter} and the Heterodyne Receiver Array CO Line Extragalactic Survey \citep[HERACLES\footnote{\url{https://www2.mpia-hd.mpg.de/HERACLES/Overview.html}};][]{2009Leroy}. The basic characteristics of those data sets are briefly described in the following.  

The THINGS survey is a high spectral ($\leqslant 5.2 \text{ km}\text{ s}^{-1}$) and spatial (beam size of $\sim6''$) resolution survey of the 21 cm line (\HI{} emission) targeting 34 nearby galaxies using the National Radio Astronomy Observatory (NRAO) Very Large Array (VLA). The primary beam (field of view) has a full width at half-maximum (FWHM) of $32'$. However, we use the ``robust'' weighted maps products which have a typical beam size of $\sim 6''$ \citep{2008Walter, 2008Bigiel}. The spatial sampling of the integrated \HI{} (moment 0) map is $1.5''\text{ pixel}^{-1}$. 
The typical $1\sigma$ rms noise achieved by the observations with average exposure of 7 hours on source (observing two polarizations) is $0.4\text{ mJy}\text{ beam}^{-1}$ (or $6$ K), corresponding to a column density limit of $\sim 3.2\times10^{20}\text{ cm}^{-2}$.

The HERACLES survey observed spatially resolved CO $J=2\rightarrow 1$ lines over the optical extent of 48 nearby galaxies using the IRAM 30-m telescope. The observation used the Heterodyne Receiver Array \citep[HERA;][]{2004Schuster} multipixel receiver. The integrated CO intensity map produced by this survey has a spatial resolution of $\sim 11''$ and spatial sampling of $2.0''\text{ pixel}^{-1}$. 
For a source with a full line width of $\sim 10\text{ km}\text{ s}^{-1}$, the typical rms sensitivity translates to a $3\sigma$ point source sensitivity of $4.8\times 10^{5}d_{10}\text{ K}\text{ km}\text{ s}^{-1}\text{pc}^{2}$, where $d_{\rm 10}$ is the distance to the source divided by 10 Mpc. This corresponds to an $\text{H}_{2}$ limiting mass of $2.7\times 10^{6} d_{10} M_{\odot}$ \citep{2009Leroy}.

\subsection{Sample Selection} \label{sec:sample_select}
The sample galaxies analyzed in this work are selected from the HERACLES survey, based on the following criteria:
\begin{enumerate}
\item Proximity, such that the galaxies are located within 20 Mpc distance. This criteria is applied to achieve a physical spatial resolution of $\sim 1-2$ kpc with the set of multiband imaging data used in this work (see Section~\ref{sec:broad_band_images}).
\item Observed by the SDSS imaging survey\footnote{The survey coverage of the SDSS DR12 is $14,555$ $\text{deg}^{2}$}. 
\item Having CO detection in more than 50\% of pixels within $3\times R_{e}$, where $R_{e}$ is the half-mass radius, which is the semi-major axis length of an ellipse that covers half of the total $M_{*}$.
\item Having a spiral (i.e.,~disky) morphology.  
\end{enumerate}

The above criteria result in 10 galaxies whose basic global properties are summarized in Table~\ref{tab:sample_galaxies}. The sources of their FUV--FIR imaging data are summarized in Table~\ref{tab:surveys_GLX_IRC_MIPS}.

There are other surveys that study the spatially resolved properties of stellar population and molecular gas in the ISM, such as the ALMA-MaNGA QUEnching and STar Formation \citep[ALMaQUEST;][]{2020Lin} and the Physics at High Angular Resolution in Nearby GalaxieS survey \citep[PHANGS;][]{2021Lee, 2021Leroy, 2021Emsellem}. The ALMaQUEST survey provides the spatially-matched optical integral field spectroscopy (IFS) data from the Mapping nearby Galaxies at Apache Point Observatory \citep[MaNGA;][]{2015Bundy} survey and the CO $J=1\rightarrow 0$ line emission maps obtained with the Atacama Large Millimeter/submillimeter Array (ALMA), while the PHANGS survey provides a combined high spatial resolution data set that consists of optical IFS data from the Multi Unit Spectroscopic Explorer (MUSE) at the Very Large Telescope (VLT), CO $J=2\rightarrow 1$ line emission maps obtained with ALMA, and optical imaging data obtained with the Hubble Space Telescope (HST). We adopt our current data sets because of the following advantages over the above surveys: (1) the multiwavelength imaging data used in our work allow us to apply the FUV--FIR SED fitting and self-consistently measure the properties of the stellar population and dust, which is not directly provided by the above surveys, and (2) the wider coverage of the galaxy region ($>3R_{e}$) provided by the imaging data used in our work. For comparison, the ALMaQUEST survey achieved a radial coverage of $\sim 1.5-2.5R_{e}$ \citep{2020Lin}, while the radial coverage of PHANGS survey varies: $\sim 1-2R_{e}$ for the CO maps and $>3R_{e}$ for the IFS data \citep{2021Emsellem}.

Among the sample galaxies, two galaxies (NGC 4254 and NGC 4579) are not covered in the THINGS survey, although they are covered in the HERACLES survey. They are kept in our sample because the available molecular gas information still provides great insights into the gas properties.  

\begin{figure}
\centering
\includegraphics[width=0.45\textwidth]{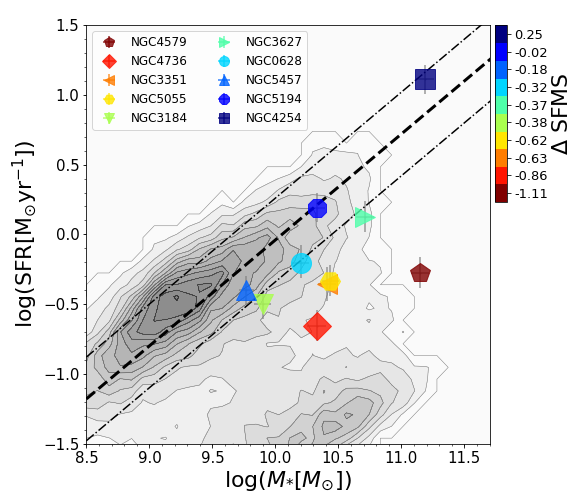}
\caption{Distribution of the sample galaxies on the $M_{*}$--SFR plane (i.e.,~SFMS plane). The background contours show the distribution of local galaxies that are taken from the MPA--JHU value added catalog. The $M_{*}$ and SFR of the sample galaxies are derived from the spatially resolved SED fitting using \texttt{piXedfit}. The black dashed line and dot--dashed lines are the SFMS ridge line from \citet{2015Renzini} and $\pm0.3$ dex around the SFMS, respectively. The sample galaxies are color-coded based on their distances from the SFMS ridge line.}
\label{fig:plt_sample_SFMS_new}
\end{figure}

Figure~\ref{fig:plt_sample_SFMS_new} shows the distribution of the sample galaxies on the stellar mass ($M_{*}$)--SFR plane (i.e.,~star-forming main sequence or SFMS plane). The contours in the background show the distribution of local galaxies within $0.001\leqslant z \leqslant 0.05$ that are taken from the MPA--JHU value added galaxy catalog\footnote{\url{https://wwwmpa.mpa-garching.mpg.de/SDSS/DR7/}} \citep{2003Kauffmann, 2004Brinchmann, 2004Tremonti, 2007Salim}. In this catalog, the $M_{*}$ and SFR are derived from fitting to the 5-band SDSS photometry and emission lines modeling, respectively. A division by a factor of $1.06$ has been applied to correct for the slight excess in the $M_{*}$ and SFR between the \citet{2001Kroupa} IMF (which is assumed in the MPA-JHU catalog) and the \citet{2003Chabrier} IMF \citep[e.g.,][]{2014Speagle}. The dashed line represents the SFMS ridge line from \citet{2015Renzini}, while the dot--dashed lines are $\pm 0.3$ dex around the SFMS ridge line. The integrated $M_{*}$ and SFR of our sample galaxies are derived by summing up the $M_{*}$ and SFR of pixels within the galaxy's region of interest (to be described in Section~\ref{sec:image_process} and~\ref{sec:sed_fitting}). The sample galaxies are color-coded based on their distances from the SFMS ridge line. The color bar is arbitrarily binned such that each color represents one galaxy. This color-coding will be used throughout this paper.

From this figure, we can see that our sample galaxies spread widely across the SFMS plane. NGC4254, NGC5194, and NGC 5457 reside on the star-forming sequence, while the other galaxies are located around the boundary between the star-forming sequence and green-valley (NGC 628, NGC 3627, and NGC 3184), and on the green-valley region (NGC 5055, NGC 3351, NGC 4736,  and NGC 4579). NGC 4254 is the most massive and actively star-forming among the galaxies in our sample. We note the possibility of discrepancies in the $M_{*}$ and SFR measurements between our method and that of the MPA-JHU team. Therefore, the background contours in the figure is only intended for reference and demonstration purposes. Investigation of the possible discrepancies between our $M_{*}$ and SFR estimates with those taken from the MPA-JHU is beyond the scope of this paper.

\section{Methodology} \label{sec:method}

\subsection{Data Analysis} \label{sec:data_analysis}
We use \verb|piXedfit| \citep{2021Abdurrouf} to spatially match (in spatial resolution and sampling) the multiband FUV--FIR imaging data, combine neighboring pixels to get the spatially resolved SEDs with sufficient signal-to-noise (S/N) ratios, and perform SED fitting on the SEDs of spatial bins to obtain the spatially resolved properties of the galaxies. In the following, we briefly describe the data anlysis carried out in this work.

\subsubsection{Processing the FUV--FIR Imaging Data} \label{sec:image_process}
We use the \verb|piXedfit_images| module in \verb|piXedfit| to perform spatial matching of the FUV--FIR imaging data. The image processing for each galaxy includes background subtraction, PSF matching, spatial resampling and reprojection, defining galaxy's region of interest, calculating fluxes of pixels, and correcting those fluxes for the foreground Galactic dust extinction. First, a variance image (i.e.,~square of the uncertainty image) associated with each science image is constructed. Whenever available, we use the uncertainty image provided by the survey from which the image is obtained. Otherwise, we generate it using a specific function in the \verb|piXedfit_images| module. Most of the archival imaging data used in this work are background free (i.e.,~the background subtration has been done). We only conduct background subtraction to the WISE images, which are not background free.

Given that none of the galaxies in our sample is contained within a single SDSS frame, we build mosaics using the \texttt{Montage} \citep{2010Jacob} software through the Python wrapper \texttt{montage\_wrapper}\footnote{\url{https://montage-wrapper.readthedocs.io/en/v0.9.5/}.}. The mosaics are made for both the science and variance images.
 
In the image processing, the multiband images are brought to the same spatial resolution (i.e.,~PSF size), sampling (i.e.,~pixel size), and projection. The final spatial resolution and sampling are defined by the largest PSF FWHM and pixel size among the imaging data. The PSF matching process, which is done by convolving images with pre-calculated kernels, uses convolution kernels from \citet{2011aniano}\footnote{\url{https://www.astro.princeton.edu/~ganiano/Kernels.html}}.

In this work, we also study the effects of varying the spatial resolution and the wavelength coverage of the photometric SED on the overall derived spatially resolved properties. To achieve this, for each galaxy, the analyses are done in two different sets of imaging data such that we get two different final spatial resolutions, spatial sampling, and wavelength coverage in the final product of the image processing. The two sets of imaging data are: (1) 24 bands of imaging data that consist of FUV, NUV, $u$, $g$, $r$, $i$, $z$, $J$, $H$, $K_{s}$, $W1$, $W2$, $W3$, $W4$, $3.6$ $\mu$m, $4.5$ $\mu$m, $5.8$ $\mu$m, $8.0$ $\mu$m, $24$ $\mu$m, $70$ $\mu$m, $100$ $\mu$m, $160$ $\mu$m, $250$ $\mu$m, and $350$ $\mu$m (hereafter SR1 data set), which results in the final spatial resolution of $24.88''$. The spatial sampling varies depending on the survey from which the \textit{Herschel} imaging data are taken. The VNGS survey has $8.0''\text{pixel}^{-1}$, while the KINGFISH survey has $10.0''\text{pixel}^{-1}$; and (2) 22 bands of imaging data that are the same as the SR1 setting except for the exclusion of the $250$ and $350$ $\mu$m (hereafter SR2 data set), which gives the final spatial resolution and sampling of $11.89''$ and $2.85''$, respectively.

\begin{longrotatetable}
\begin{deluxetable*}{ccccccccccccccc}
\tablenum{2}
\tablecaption{Basic Information of the Sample Galaxies\label{tab:sample_galaxies}}
\tablewidth{0pt}
\tablehead{
\colhead{Galaxy} & \colhead{Hubble type} & \colhead{Distance} & \colhead{Nuc. type} & \colhead{SFR} & \colhead{$\log(M_{*})$} & \colhead{$\log(M_{\rm dust})$} & \colhead{$\log(M_{\rm HI})$} & \colhead{$\log(M_{\text{H}_{2}})$} & \colhead{$R_{e}$} & \colhead{PA} & \colhead{$e$} & \colhead{$N_{\rm bin,1}$} & \colhead{$N_{\rm bin,2}$} & \colhead{SR1} \\
\colhead{} & \colhead{} & \colhead{(Mpc)} & \colhead{} & \colhead{($M_{\odot}\text{yr}^{-1}$)} & \colhead{($\log M_{\odot}$)} & \colhead{($\log M_{\odot}$)} & \colhead{($\log M_{\odot}$)}  & \colhead{($\log M_{\odot}$)} & \colhead{(kpc)} & \colhead{} & \colhead{} & \colhead{} & \colhead{} & \colhead{(kpc)} \\
\colhead{(1)} & \colhead{(2)} & \colhead{(3)} & \colhead{(4)} & \colhead{(5)} & \colhead{(6)} & \colhead{(7)} & \colhead{(8)} & \colhead{(9)} & \colhead{(10)} & \colhead{(11)} & \colhead{(12)} & \colhead{(13)} & \colhead{(14)} & \colhead{(15)}
}
\startdata
NGC 0628 & SAc & $9.77$ & \nodata & $0.62_{-0.11}^{+0.15}$ & $10.20_{-0.07}^{+0.07}$ & $7.75_{-0.07}^{+0.07}$ &  $9.58$  & $9.42\pm 0.06$ & $3.23$ & $18.01$ & $0.11$ & 84 & 265 & $1.13$ \\
NGC 3184 & SABcd & $12.27$ & non-AGN & $0.32_{-0.11}^{+0.11}$ & $9.90_{-0.07}^{+0.07}$ & $7.52_{-0.07}^{+0.07}$ &  $9.19$ & $9.08\pm 0.08$ & $2.98$ & $88.88$ & $0.04$ & 54 & 259 & $1.02$\\
NGC 3351 & SBb & $10.50$ & non-AGN & $0.44_{-0.12}^{+0.12}$ & $10.41_{-0.07}^{+0.07}$ & $7.64_{-0.07}^{+0.08}$ &  $9.18$ & $9.25\pm 0.07$ & $2.49$ & $81.06$ & $0.07$ & 75 & 159 & $1.34$ \\
NGC 3627 & SABb & $9.04$ & AGN & $1.34_{-0.12}^{+0.13}$ & $10.71_{-0.07}^{+0.07}$ & $7.88_{-0.07}^{+0.07}$ &  $9.14$ & $9.68\pm 0.06$ & $2.89$ & $76.38$ & $0.25$ & 73 & 223 & $1.25$ \\
NGC 4254 & SAc & $15.19$ & \nodata & $12.91_{-0.30}^{+0.30}$ & $11.18_{-0.07}^{+0.07}$ & $8.71_{-0.07}^{+0.07}$ & \nodata & $10.68\pm 0.05$ & $8.83$ & $147.84$ & $0.26$ & 38 & 192 & $1.80$ \\
NGC 4579 & SABb & $18.40$ & AGN & $0.53_{-0.12}^{+0.12}$ & $11.15_{-0.07}^{+0.07}$ & $8.06_{-0.07}^{+0.07}$ & \nodata & $9.72\pm 0.06$  & $5.06$ & $167.02$ & $0.28$ & 31 & 89 & $2.18$ \\
NGC 4736 & SAab & $4.59$ & AGN & $0.22_{-0.11}^{+0.11}$ & $10.33_{-0.07}^{+0.07}$ & $7.23_{-0.07}^{+0.08}$ &  $8.65$ & $8.85\pm 0.11$ & $0.68$ & $9.73$ & $0.28$ & 85 & 279 & $0.53$ \\
NGC 5055 & SAbc & $9.04$ & AGN & $0.46_{-0.11}^{+0.11}$ & $10.44_{-0.07}^{+0.07} $ & $7.91_{-0.07}^{+0.07}$ &  $9.44$ & $9.50\pm 0.07$ & $3.14$ & $9.89$ & $0.54$ & 135 & 393 & $0.83$ \\
NGC 5194 & SABbc & $8.58$ & AGN &$1.55_{-0.11}^{+0.11}$ & $10.33_{-0.07}^{+0.07}$ & $7.72_{-0.07}^{+0.07}$ &  $9.14$ & $9.59\pm 0.07$ & $3.24$ & $123.58$ & $0.33$ & 218 & 709 & $0.80$ \\
NGC 5457 & SABcd & $	6.95$ & non-AGN & $0.40_{-0.11}^{+0.11}$ & $9.77_{-0.07}^{+0.07}$ & $7.33_{-0.07}^{+0.07}$ &  $9.09$ & $8.89\pm 0.13$ & $2.68$ & $140.00$ & $0.20$ & 256 & 1037 & $0.42$ \\
\enddata
\tablecomments{
Col. (1): Galaxy name. 
Col. (2): Morphological type in the Hubble sequence which are taken from \citet{2003Kennicutt, 2011Kennicutt}. 
Col. (3): Distance to the galaxy which are adopted from the NASA/IPAC Extragalactic Database (NED) website. NGC 628: based on tip of the red giant branch (TRGB) method carried by \citet{2017McQuinn}. NGC 3184, NGC 4254, NGC 4579: based on the average of distances obtained with various methods as compiled by the NED. NGC 3351, NGC 3627, NGC 4736, NGC 5055, and 5457: based on the average of distances obtained with the combination of several methods, including cepheids, TRGB, Type Ia supernova (SNIa), Tully-Fisher relation, and the surface brightness fluctuation (SBF) \citep{2013Tully}. NGC 5194: based on the TRGB method as performed by \citep{2016McQuinn}.
Col. (4): Nuclear type information adopted from \citet{2011Kennicutt}, which is based on \citet{2010Moustakas} and \citet{1997Ho}. 
Col. (5): Integrated SFR that is obtained from summing up the spatially resolved SFR derived using \texttt{piXedfit}.
Col. (6): Integrated stellar mass that is obtained from summing up the spatially resolved $M_{*}$ derived using \texttt{piXedfit}. 
Col. (7): Integrated dust mass from \texttt{piXedfit}.
Col. (8): Integrated neutral gas mass derived from the $\Sigma_{\rm HI}$ map. 
Col. (9): Integrated molecular gas mass derived from the $\Sigma_{\rm H_{2}}$ map.
Col. (10): Half-mass radius, which is the semi-major axis length of an ellipse that covers half of the total stellar mass.
Col. (11): Position angle of the elliptical aperture, which is defined as an angle between the semi-major axis and the positive $X$-axis. The angle is measured rotating towards the positive $Y$-axis.
Col. (12): Ellipticity of the elliptical aperture.  
Col. (13): Number of bins in the SR1 data set.  
Col. (14): Number of bins in the SR2 data set.
Col. (15): Physical spatial resolution (corresponding to the PSF FWHM) of the SR1 data set.
}
\end{deluxetable*}
\end{longrotatetable}

\begin{deluxetable}{ccccc}
\tablenum{3}
\tablecaption{Sources of the Imaging Data in the GALEX, \textit{Spitzer} (IRAC and MIPS), and \textit{Herschel} (PACS and SPIRE) bands \label{tab:surveys_GLX_IRC_MIPS}}
\tablewidth{0pt}
\tablehead{
\colhead{Galaxy} & \multicolumn{4}{c}{Survey Name} \\
\cline{2-5}
\colhead{} & \colhead{GALEX} & \colhead{IRAC} & \colhead{MIPS} & \colhead{\textit{Herschel}}
}
\decimals
\startdata
NGC 0628 & GII & SINGS & SINGS & KINGFISH\\
NGC 3184 & AIS & SINGS & SINGS & KINGFISH\\
NGC 3351 & NGS & SINGS & SINGS & KINGFISH\\
NGC 3627 & NGS & SINGS & SINGS & KINGFISH\\
NGC 4254 & GII & SINGS & SINGS & KINGFISH\\
NGC 4579 & NGS & SINGS & SINGS & KINGFISH\\
NGC 4736 & NGS & SINGS & SINGS & KINGFISH\\
NGC 5055 & NGS & SINGS & SINGS & KINGFISH\\
NGC 5194 & GII & SINGS & SINGS & VNGS\\
NGC 5457 & NGS & SEIP & LVL & KINGFISH
\enddata
\end{deluxetable}

After the spatial matching of the multiband imaging data, the next step is defining the region of interest\footnote{A region of interest within which the spatially resolved SED fitting will be performed.} associated with each galaxy. Briefly speaking, this process is done by first constructing a segmentation map in each band using the \texttt{SExtractor} \citep{1996bertin}, and then merging all the segmentation maps. The merged segmentation map is then used as an initial definition for the galaxy's ``region''. After that, the muliband fluxes and flux uncertainties of pixels within this region are calculated. Finally, we crop regions associated with bright foreground stars. This process produces a total of 20 data cubes (10 for each of the SR1 and SR2 data formats) which contain the pixel-wise SEDs of the galaxies. 

\begin{figure*}
\centering
\includegraphics[width=0.9\textwidth]{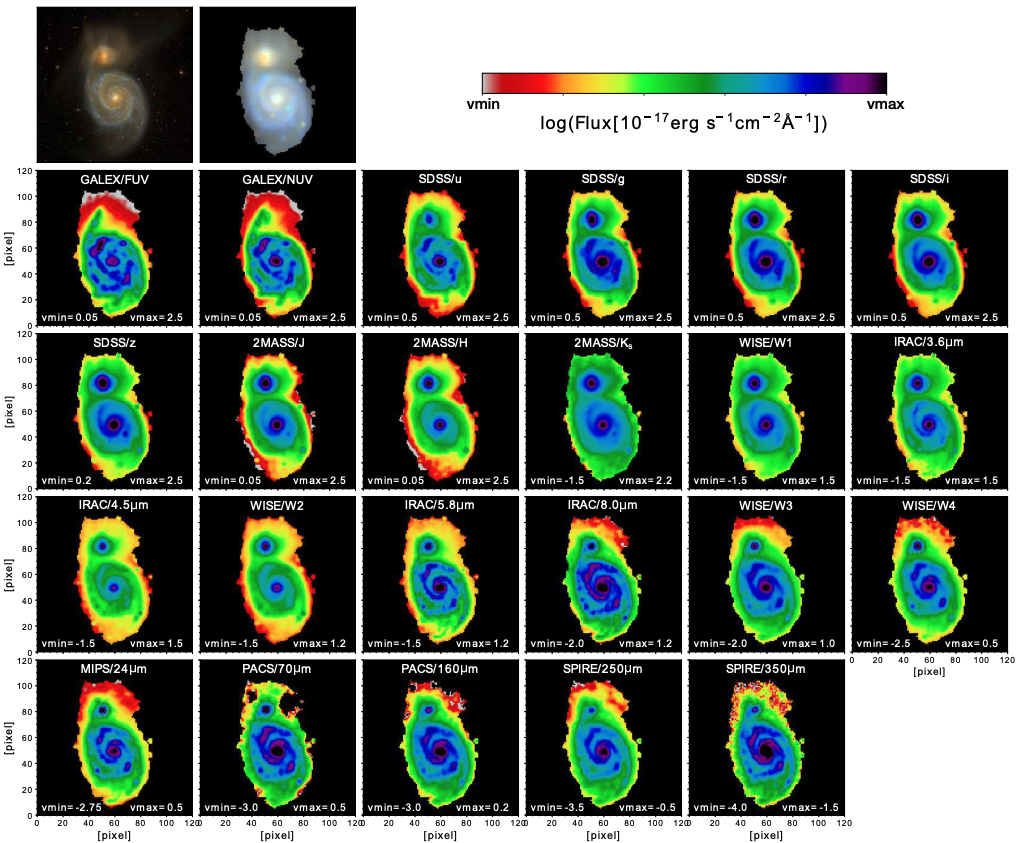}
\caption{Example of the maps of multiband fluxes of NGC 5194 that are obtained from the image processing on the 23-band imaging data that range from the GALEX/FUV to \textit{Herschel}/SPIRE $350$ $\mu$m (i.e.,~the SR1 data format). The two panels in the first row show $gri$ composite images: one in the original SDSS spatial resolution (left panel) and the other in the spatial resolution of SPIRE $350$ $\mu$m (right panel), which is the final spatial resolution achieved in the image processing. Different ranges of fluxes are applied to the colorbars of different maps to enhance the spiral arms. The minimum (vmin) and maximum (vmax) values of the fluxes (in the same units as that of the colorbars) are shown in each map.}
\label{fig:maps_fluxes_SR1_new1_ed}
\end{figure*}

Figure~\ref{fig:maps_fluxes_SR1_new1_ed} shows the maps of multiband fluxes of NGC 5194 (including NGC 5195) obtained from the image processing on the SR1 data set. The two panels in the first row show $gri$ composite images, one in the original SDSS spatial resolution (left panel) and the other in the spatial resolution of SPIRE $350$ $\mu$m (right panel), which is the final spatial resolution of the reduced 3D data cube after the image processing. Different ranges of flux are applied to the colorbars of different maps to enhance the spiral arms feature. The minimum (vmin) and maximum (vmax) values of the fluxes are shown in each map.

\subsubsection{Spatial Matching with the \HI{} and CO Data} \label{sec:matching_HICO}
For a self-consistent analysis of the spatially resolved properties and distributions of the stellar, dust, and gas components, we spatially match the \HI{} and CO maps with the reduced FUV--FIR imaging data cubes (described in Section~\ref{sec:image_process}). 

First, we convert the pixel values of the \HI{} and CO images into the neutral gas mass and molecular gas mass surface densities ($\Sigma_{\rm HI}$ and $\Sigma_{\rm H_{2}}$, respectively) in units of $M_{\odot}\text{ kpc}^{-2}$. To convert the integrated \HI{} intensity (in units of $\text{Jy}\text{ beam}^{-1}\text{ m}\text{ s}^{-1}$) into the $\Sigma_{\rm HI}$, we use the following equation adapted from \citet{2008Walter}
\begin{equation}
	\begin{aligned}
	&	\Sigma_{\rm HI} [M_{\odot}\text{kpc}^{-2}] \\
	& = \frac{8.86\times 10^{6} \times I_{\rm HI}[\text{Jy }\text{beam}^{-1}\text{m }\text{s}^{-1}]}{\text{FWHM}_{\rm maj}[''] \times \text{FWHM}_{\rm min}['']},
	\end{aligned}
\end{equation}   
where the $\text{FWHM}_{\rm maj}$ and $\text{FWHM}_{\rm min}$ are the major and minor axes of the beam in arcsec which are taken from Table 3 of \citet{2008Walter}. To convert the CO line intensity ($I_{\text{CO } J=2\rightarrow 1}$; in units of $\text{K}\text{ km}\text{ s}^{-1}$) map to $\Sigma_{\rm H_{2}}$, we use the following equation
\begin{equation}
	\begin{aligned}
	& \Sigma_{\rm H_{2}} [M_{\odot}\text{kpc}^{-2}] \\
	& = \alpha_{\rm CO} 	\frac{I_{\text{CO } J=2\rightarrow 1}[\text{K }\text{km }\text{s}^{-1}]}{R_{\rm 21}},
	\end{aligned}
\end{equation}
where $\alpha_{\rm CO}$ is the CO-to-$\text{H}_{2}$ conversion factor. In the analysis of this paper, we assume a constant $\alpha_{\rm CO}$ as $4.35\times 10^{6}$ $M_{\odot} \text{kpc}^{-2} (\text{K }\text{km }\text{s}^{-1})^{-1}$, which is the typical value observed in the Milky Way \citep[e.g.,][]{1987Solomon, 2013Bolatto}. Since the standard $\alpha_{\rm CO}$ is quoted for $I_{\text{CO } J=1\rightarrow 0}$, we divide the $I_{\text{CO } J=2\rightarrow 1}$ with a fixed line ratio $R_{21}=(2\rightarrow 1)/(1\rightarrow 0)$ of $0.8$ \citep{2009Leroy}. For later analysis, we also calculate the total gas mass surface density as $\Sigma_{\rm gas}=\Sigma_{\rm atom}+\Sigma_{\rm H_{2}}$ where $\Sigma_{\rm atom}=1.36\times \Sigma_{\rm HI}$ is the surface density of atomic gas including the contribution of helium \citep[e.g.,~][]{2013Sandstrom}.

The \HI{} and $\text{H}_{2}$ maps are next subject to the same PSF matching, resampling and projection steps as outlined in Section 3.1.1. Here, we assume the PSFs to be Gaussian with FWHMs of $6.0''$ and $11.0''$ for \HI{} and CO images, respectively. After that, we crop the \HI{} and $\text{H}_{2}$ maps based on the galaxy's regions of interest defined previously (see Section~\ref{sec:image_process}).

\subsubsection{Pixel Binning} \label{sec:pixel_binning}
In order to obtain spatially resolved SEDs with sufficient S/N ratio in all bands, we perform a pixel binning to each data cube using the \verb|piXedfit_bin| module. Detailed explanation of the procedures adopted in the pixel binning process is given in \citet[][Section 3.3.~therein]{2021Abdurrouf}. In short, the \verb|piXedfit_bin| module is capable of binning neighboring pixels that have similarity in SED shape (within a certain chi-square limit, $\chi_{\text{max,bin}}^{2}$) and reaching target S/N thresholds that are set in all bands. Briefly speaking, the pixel binning scheme is performed by growing the size of bins (which is started from a brightest pixel) and including more neighboring pixels until the S/N thresholds in all bands are achieved.

We perform the pixel binning with the following requirements: (1) reduced $\chi_{\text{max,bin}}^{2}$ of $5.0$ in the evaluation of the SED shape similarity, (2) the S/N threshold of $5$ in all bands, and (3) minimum diamater of a bin ($D_{\text{min,bin}}$) of 4 and 8 pixels for the SR1 and SR2 data cubes, respectively. These $D_{\text{min,bin}}$ values are selected because they are similar to the PSF FWHM of the data cubes. Besides avoiding picking a single bright pixel (that meets the S/N thresholds requirements) as a single bin, this $D_{\text{min,bin}}$ parameter also has an important benefit in terms of fitting the SED of the central regions of an AGN-host galaxy. Since the morphology of the central bright AGN component follows the PSF shape, it is expected that the significant contribution of the AGN is enclosed within the central spatial bin. Thus, we only need to include the AGN component in the modeling and fitting of the SED of the central spatial bin. Each galaxy has a different number of spatial bins in the SR1 and SR2 data cubes (see Table~\ref{tab:sample_galaxies}).

\subsection{SED Fitting Analysis} \label{sec:sed_fitting}

After the pixel binning, SED fitting is performed on all the spatial bins for each galaxy using the \verb|piXedfit_fitting| module. Detailed descriptions of the adopted methods in the SED modeling and fitting in the \verb|piXedfit_fitting| module are presented in \citet[][Sections 3.4 and 4 therein]{2021Abdurrouf}. Briefly speaking, the SED modeling uses the Flexible Stellar Population Synthesis \citep[\texttt{FSPS}\footnote{\url{https://github.com/cconroy20/fsps}};][]{2009Conroy,2010Conroy} code through the \verb|python-fsps|\footnote{\url{http://dfm.io/python-fsps/current/}} package \citep{2014Foreman}. The SED modeling incorporates four main emission sources: stellar, nebular, dust, and AGN dusty torus. The nebular emission modeling uses the \texttt{CLOUDY} code \citep{1998Ferland, 2013Ferland} which was implemented in the \texttt{FSPS} by \citet{2017Byler}. 
We fix the ionization parameter ($U$) in the nebular emission modeling to $0.01$, which is the default value in the FSPS. This is because the ionization parameter has only a small influence on the luminosities of bright emission lines (e.g.,~$\text{H}_{\alpha}$ and $\text{H}_{\beta}$) and contribution of emission lines to the broad-band photometry of the local galaxies are not significant \citep[see e.g.,][]{2017Leja}. Therefore, we assume a fixed $U$ for simplicity.  

The dust emission modeling uses the \citet{2007Draine} templates, while the modeling of the AGN dusty torus emission uses the \citet{2008Nenkova_a, 2008Nenkova_b} \texttt{CLUMPY} models. Detailed descriptions on the implementations of the dust emission and AGN dusty torus emission in the FSPS are given in \citet{2017Leja, 2018Leja}. We only include the AGN component in the fitting for the central spatial bin of the galaxies that host an AGN and those that are unclassified (see Table~\ref{tab:sample_galaxies}). In the analysis throughout this work, we assume the \citet{2003Chabrier} IMF, Padova isochrones \citep{2000Girardi,2007Marigo,2008Marigo}, MILES stellar spectral library \citep{2006Sanchez-Blazquez,2011Falcon}, and double power-law star formation history (SFH). It has been shown in \citet{2021Abdurrouf} that this SFH form can give good inferences of the stellar population properties as well as the SFH of galaxies, as tested using the mock SEDs of the IllustrisTNG galaxies. For simulating dust attenuation in the model SEDs, we use the two-component dust attenuation law of \citet{2000Charlot}. In addition to the attenuation due to the diffuse ISM, this dust attenuation law yields an extra attenuation to young stars ($t<10$ Myr) which reside in the dense molecular clouds.    

Overall, the SED fitting adopts the Bayesian inference technique with two kinds of posterior sampling methods: the Markov Chain Monte Carlo (MCMC) and the random dense sampling of parameter space \citep[RDSPS;][]{2021Abdurrouf}. In this work, we use the MCMC method and implement it via the \verb|emcee|\footnote{\url{https://github.com/dfm/emcee}} package \citep{2013Foreman, 2018Foreman_zenodo, 2019Foreman}. We use this method for the following two reasons: (1) the MCMC method is expected to be able to effectively sample the posterior probability distributions of the parameters when many are involved in the fitting, such as that carried out in our work; and (2) we do not analyze a large number of sample galaxies in this work.

A list of the free parameters in the fitting process, along with their description and the assumed priors are given in Table~\ref{tab:free_params}. Although the dust mass is not a free parameter in the dust emission modeling, it can be retrieved from the FSPS\footnote{Using the property \texttt{dust\_mass} in the \texttt{fsps.StellarPopulation} class.}. In principle, the dust mass can be calculated for a given dust emission SED with a certain set of parameters ($U_{\rm min}$, $Q_{\rm PAH}$, and $\gamma_{e}$; see Section 9.5. in \citealt{2007Draine}). In total, we perform SED fitting to 4654 spatial bins, which are the sum of the SR1 (1049) and SR2 (3605) data cubes.  

\begin{deluxetable*}{lp{8cm}p{5cm}l}
\tablenum{4}
\tablecaption{Description of the Free Parameters in the SED Fitting and the Assumed Priors. \label{tab:free_params}}
\tablewidth{0pt}
\tablehead{
\colhead{Parameter} & \colhead{Description} & \colhead{Prior} & \colhead{Sampling/Scale}
}
\startdata
$M_{*}$ & Stellar mass & Uniform: min$=\log(s_{\rm best})-2$, max$=\log(s_{\rm best})+2$\textsuperscript{a} & Logarithmic \\
$Z_{*}$ & Stellar metallicity & Uniform: min$=-2.0+\log(Z_{\odot})$, max$=0.2+\log(Z_{\odot})$ & Logarithmic \\
$t$ & Time since the onset of star formation ($\text{age}_{\rm sys}$) \textsuperscript{b} & Uniform: min$=-1.0$, max$=1.14$ & Logarithmic \\
$\tau$ & Parameter that controls the peak time in the double power-law SFH\textsuperscript{b} & Uniform: min$=-1.5$, max$=1.14$ & Logarithmic \\
$\alpha_{\rm SFH}$ & Parameter in the double power-law SFH that controls the slope of the falling star formation episode\textsuperscript{b} & Uniform: min$=-2.0$, max$=2.0$ & Logarithmic \\
$\beta_{\rm SFH}$ & Parameter in the double power-law SFH that controls the slope of the rising star formation episode\textsuperscript{b} & Uniform: min$=-2.0$, max$=2.0$ & Logarithmic \\
$\hat{\tau}_{1}$ & Dust optical depth of the birth cloud in the \citet{2000Charlot} dust attenuation law & Uniform: min$=0.0$, max$=3.0$ & Linear \\
$\hat{\tau}_{2}$ & Dust optical depth of the diffuse ISM in the \citet{2000Charlot} dust attenuation law & Uniform: min$=0.0$, max$=3.0$ & Linear \\
$n$ & Power law index in the dust atttenuation curve for the diffuse ISM in the \citet{2000Charlot} dust attenuation law & Uniform: min$=-2.2$, max$=0.4$ & Linear \\
$U_{\rm min}$ & Minimum starlight intensity that illuminate the dust & Uniform: min$=-1.0$, max$=1.176$ & Logarithmic \\
$\gamma_{e}$ & Relative fraction of dust heated at a radiation field strength of $U_{\rm min}$ and dust heated at $U_{\rm min}<U\leq U_{\rm max}$ & Uniform: min$=-3.0$, max$=-0.824$ & Logarithmic \\
$Q_{\rm PAH}$& Fraction in percent (\%) of total dust mass that is in the polycyclic aromatic hydrocarbons (PAHs) & Uniform: min$=-1.0$, max$=0.845$ & Logarithmic \\
$f_{\rm AGN,*}$\textsuperscript{c} & AGN luminosity as a fraction of the bolometric stellar luminosity & Uniform: min$=-5.0$, max$=0.48$ & Logarithmic \\
$\tau_{\rm AGN}$&Optical depth of the AGN dusty torus & Uniform: min$=0.70$, max$=2.18$ & Logarithmic
\enddata
\tablecomments{
\textsuperscript{a} $s_{\rm best}$ is the normalization of model SED that is obtained from the initial fitting with the $\chi^{2}$ minimization method (see Section 4.2.1 in \citealt{2021Abdurrouf}). 
\textsuperscript{b} The mathematical form of the double power-law SFH is given in \citet[][Equation 7 therein]{2021Abdurrouf}. We change the symbols for the rising and falling slopes of the double power-law SFH in this paper to $\beta_{\rm SFH}$ and $\alpha_{\rm SFH}$ from formerly $\beta$ and $\alpha$ in \citet{2021Abdurrouf} to avoid confusion with the UV slope $\beta$ that will be discuss in Section~\ref{sec:IRX_beta}.   
\textsuperscript{c} $f_{\rm AGN, *}$ is a native parameter in the FSPS which determines the relative amount of the AGN luminosity with respect to the bolometric stellar luminosity, so it can be greater than one. 
}
\end{deluxetable*} 

\begin{figure*}
\centering
\includegraphics[width=1.0\textwidth]{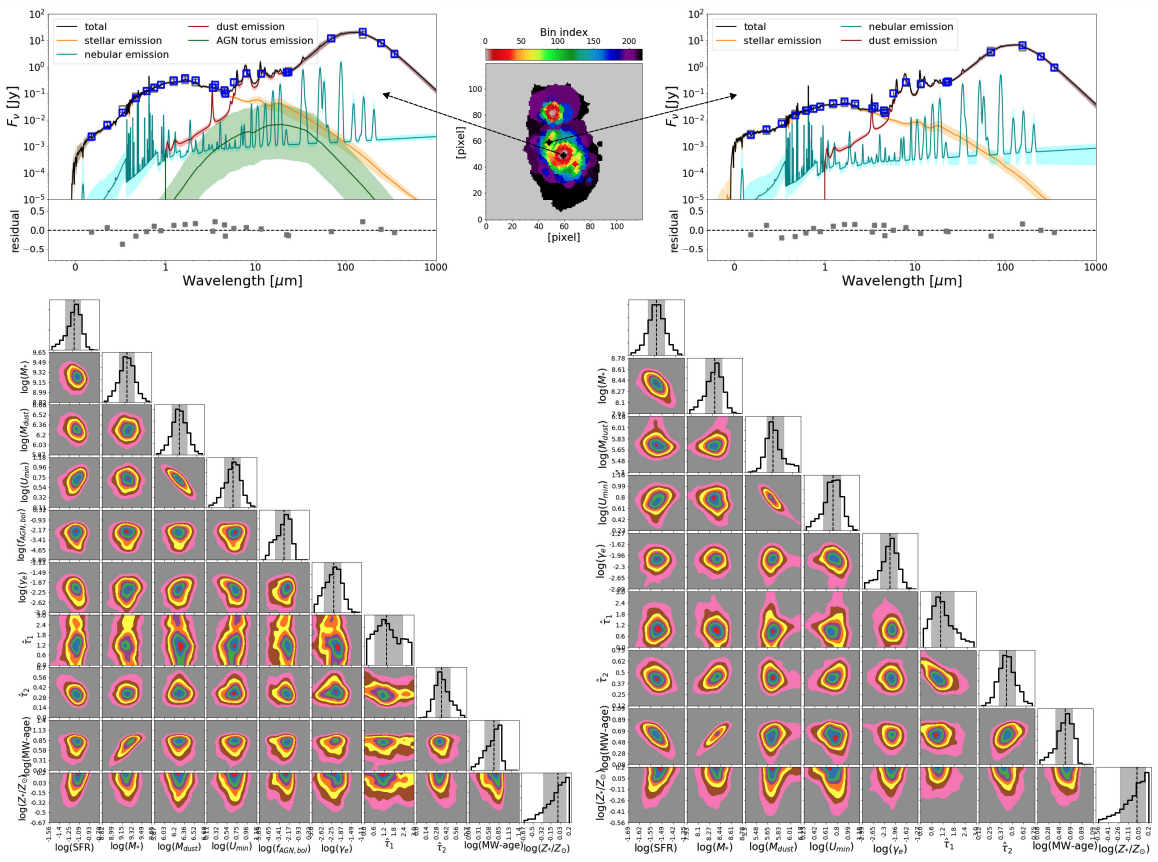}
\caption{Examples of the fitting results of two spatial bins in NGC 5194, one located in the center of the galaxy (left side) and the other in a spiral arm (right side). The map in the middle panel of the first row shows the pixel binning map, where the color bar reflects the bin index. The SED plots show the observed SEDs of the spatial bins (blue squares) and median posterior model SEDs that are broken down into their components, including the stellar (orange color), nebular (cyan color), dust (red color), and AGN dusty torus (green color) emission. The corner plots show the posterior probability distributions of some of the parameters (both the free ones as well as the dependent ones).}
\label{fig:demo_fitresult_M51}
\end{figure*}

Figure~\ref{fig:demo_fitresult_M51} shows examples of the fitting results of two spatial bins of NGC 5194 in the SR1 data format. One located in the center of the galaxy (left side) and the other in a spiral arm (right side). The pixel binning map is shown in the middle panel of the first row. The SED plots show the observed SEDs of the spatial bins (blue squares) and the median posterior model SEDs that are broken down into their components, including the stellar (orange color), nebular (cyan color), dust (red color), and AGN dusty torus (green color) emission. We include the AGN component in the fitting of the central spatial bin because this galaxy likely hosts an AGN (see Table~\ref{tab:sample_galaxies}). The corner plots show the posterior probability distributions of some of the parameters (both the free ones as well as the dependent ones). The black dashed lines and gray shaded area in the 1D histograms show the median posteriors and the uncertainties that are defined as the area between the 16th and 84th percentiles. 

In contrast to some other SED fitting codes that define SFR as an average over a certain timescale (e.g.,~the last $100$ Myr) from the SFH, the SFR in \texttt{piXedfit} is defined as an instantaneous rate of star formation. However, the two approaches only give a small difference ($1.2\%$) in the SFR estimate from the SED fitting that we perform here.

Detailed descriptions and demonstrations of the performances of \verb|piXedfit| in terms of the inferences of parameters are presented in \citet{2021Abdurrouf}. To further test its performance in inferring parameters of the stellar population and dust, we perform a fitting test with semi-empirical mock SEDs, which are presented in Appendix~\ref{sec:SEDfit_test_mock}. Overall, the results show that \verb|piXedfit| can recover the parameters associated with the stellar emission and dust emission quite well.    

\section{Results} \label{sec:results}

\subsection{Maps of the Spatially Resolved Properties} \label{sec:maps_properties}

In this section, we provide an overview of the maps of the spatially resolved properties of stellar populations and dust derived from the SED fitting and the gas mass surface densities ($\Sigma_{\rm HI}$ and $\Sigma_{\rm H_{2}}$). For derived parameters that scale with the flux, we divide the values of those parameters associated with a bin into the pixels that belong to the bin by weighting (i.e.~scaling) with the flux in a certain band. These parameters include $M_{*}$, SFR, and dust mass ($M_{\rm dust}$), which are scaled linearly with the fluxes of pixels in the $K_{s}$, $24$ $\mu$m, and $100$ $\mu$m bands, respectively. In selecting these reference bands, we have checked the scaling relations (in logarithmic scale) between those parameters and the fluxes in various bands and found that those reference bands result in the tightest scaling relations with slopes that are close to unity (i.e.,~linear). This further division of the bin values into pixels enhances the spatial resolution, strengthening several aspects of our analysis, including the derivation of the radial profiles, pixel-by-pixel comparisons of the surface densities of the stellar mass ($\Sigma_{*}$), SFR ($\Sigma_{\rm SFR}$), and dust mass ($\Sigma_{\rm dust}$) with the gas mass ($\Sigma_{\rm HI}$ and $\Sigma_{\rm H_{2}}$) which are already given in pixel scale. For the other parameters, their values are not divided into the pixels (i.e.,~all pixels belong to a bin have the same value).

\begin{figure*}
\centering
\includegraphics[width=1.0\textwidth]{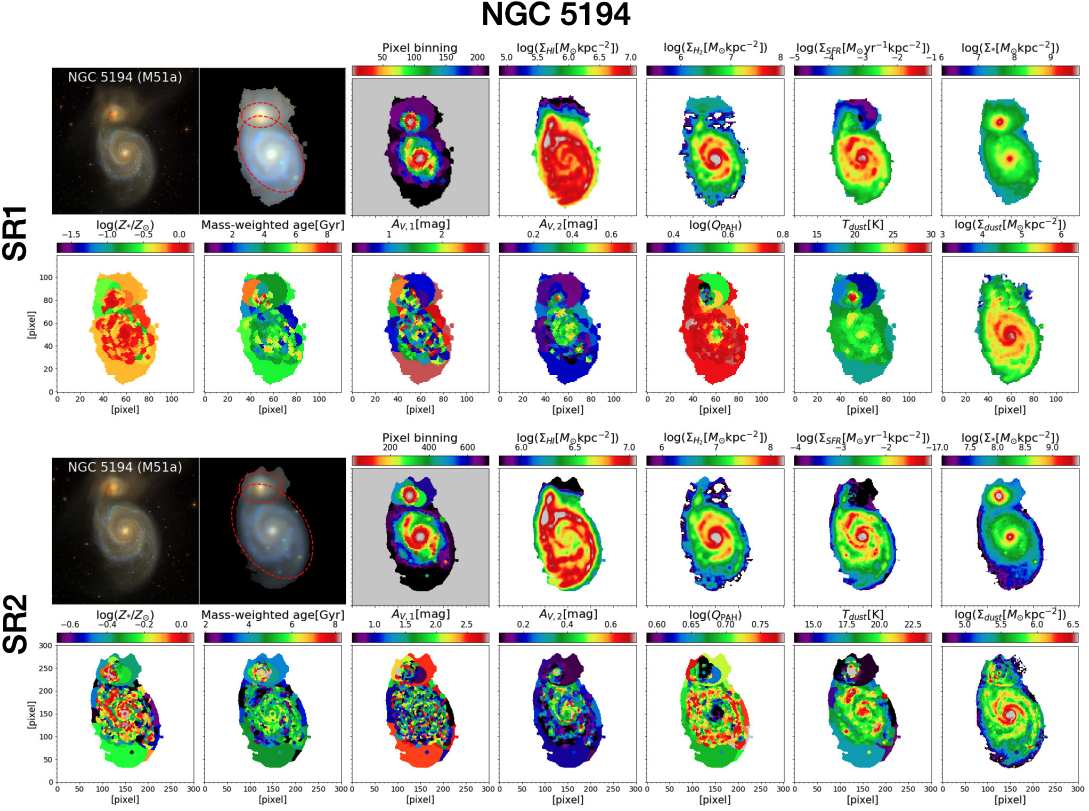}
\caption{Example of the maps of the stellar, dust, and gas properties of NGC 5194. The maps derived from the analysis with the SR1 and SR2 data sets are shown in the top two and the bottom two rows, respectively. The $gri$ composite images (in the first row of each data set) are shown in the spatial resolution of the SDSS bands (the first panel from the left) and the longest wavelength band in the data set (i.e.,~SPIRE $350$ $\mu$m in case of the SR1 data set and PACS $160$ $\mu$m in case of the SR2 data set; the second panel from the left). The maps of our entire sample (10 galaxies) are available in the online journal.}
\label{fig:plot_mapprops}
\end{figure*}

Figure~\ref{fig:plot_mapprops} shows an example of the maps of properties of NGC 5194. The top two rows show the maps from the analysis of the SR1 data set, while the bottom two rows show the maps for the SR2 data set. For each data set, in the first row from left to right we show the $gri$ composite image in the native spatial resolution of the SDSS\footnote{Obtained from \url{http://skyserver.sdss.org/dr16/en/tools/chart/image.aspx}}, $gri$ composite image in the final spatial resolution after the image processing, and maps of pixel binning, $\Sigma_{\rm HI}$, $\Sigma_{\rm H_{2}}$, $\Sigma_{\rm SFR}$, and $\Sigma_{*}$. In the second row, the panels from left to right show maps of the stellar metallicity ($Z_{*}$), mass-weighted age, dust attenuation associated with the birth clouds ($\rm{A}_{\rm V,1}$), dust attenuation associated with the diffuse ISM ($\rm{A}_{\rm V,2}$), fraction (in $\%$) of the total dust mass that is in the polycyclic aromatic hydrocarbons ($Q_{\rm PAH}$), dust temperature ($T_{\rm dust}$), and dust mass surface density ($\Sigma_{\rm dust}$).

The ellipse on the $gri$ composite images represent a radius (along the semi-major axis) of $r=3\times R_{e}$. The $R_{e}$, ellipticities ($e$), and position angles (PA) of the sample galaxies are given in Table~\ref{tab:sample_galaxies}. The $e$ and PA of the galaxies are obtained by an elliptical fitting analysis to the SPIRE $350$ $\mu$m band image using the \verb|Ellipse| function in \verb|Photutils|. This function fits elliptical isophotes of a galaxy and gives $e$ and PA at grids of radius. Because we only use the result of elliptical fitting ($e$ and PA) for the calculation of radial profiles (to be described in Section~\ref{sec:radial_profiles}), we calculate the average $e$ and PA within the intermediate radial distances (to avoid isophotes of the bulge and less robust elliptical fitting in the outskirt region) and obtain a single $e$ and PA for each galaxy. In the analysis throughout this paper, we only consider regions (i.e.,~pixels and spatial bins) that lie within $3R_{e}$. For NGC 5194, we exclude regions in the intersection with NGC 5195 (see Figure~\ref{fig:plot_mapprops}).

The quantities $\rm{A}_{V,1}$ and $\rm{A}_{V,2}$ are derived from $\hat{\tau}_{1}$ and $\hat{\tau}_{2}$, respectively, by multiplying them with a factor of $1.086$. Dust temperature $T_{\rm dust}$ is derived from $U_{\rm min}$ and $\gamma_{e}$ using the following equation, adapted from \citet{2019Utomo}
\begin{equation} \label{eq:Tdust1}
\log \langle U\rangle = (4.0+\beta_{d})\log \left(\frac{T_{\rm dust}}{18 \rm{K}} \right),
\end{equation}
which is derived based on the assumption that the dust is in a thermal equilibrium with the local radiation field and has a power-law mass absorption coefficient. The normalization is set such that $T_{\rm dust}=18$ K corresponds to the solar neighborhood radiation field, $U=1$ \citep{2014Draine}. In the above equation, $\beta_{d}$ represents the dust emissivity index, which we assume to be $1.8$ \citep[e.g.,][]{2001Dunne,2007Draine_b,2010Clements}, and $\langle U\rangle$ is calculated via (\citet{2007Draine})
\begin{equation} \label{eq:Tdust2}
\langle U \rangle = (1-\gamma_{e})U_{\rm min} + \frac{\gamma_{e} U_{\rm min} \ln(U_{\rm max}/U_{\rm min})}{1 - U_{\rm min}/U_{\rm max}}.
\end{equation}
We have assumed $U_{\rm max}=10^{6}$, which is found to reproduce the MIR photometry of the SINGS sample quite well \citep{2007Draine}.

Some general trends that can be seen from the maps are as follows: (a) $\text{H}_{2}$ tends to be more centrally-concentrated, while \HI{} tends to be more spatially extended compared with the other quantities; (b) $\Sigma_{\rm{H}_{2}}$ and $\Sigma_{\rm dust}$ tend to be higher in the spiral arms than in the inter-arm regions, indicated by the reconstructed spiral arms structure in these maps; (c) the pixel-wise $\Sigma_{\rm SFR}$ maps clearly manifest the spiral arm structures where regions with higer SFR are located; 
and (d) the central region of each galaxy has the highest dust temperature. We see a noticeably high dust temperature of $\sim 30$ K in the centers of NGC 5195 and NGC 3351. \citet{2012MentuchCooper} also observed such a high temperture ($\sim 30$ K) of dust in the central region of NGC 5195, in agreement with our result.

\subsection{Testing the empirical SFR estimate} \label{sec:comp_SFR_UVIR}

Taking advantage of the spatially resolved multiwavelength imaging data (which allows us to obtain the UV and total IR luminosities), we compare the SFR derived by \verb|piXedfit| and that obtained from an often used prescription that combines the UV and total IR luminosities. The latter is considered to be robust because of its incorporation of both dust-obscured and unobscured star formations. However, as a simple prescription, it does not consider galaxy-to-galaxy variations in the underlying stellar population properties, which are accounted for directly in SED fitting. 

We calculate the total UV and IR luminosities ($L_{\rm UV}$ and $L_{\rm IR}$, respectively) for each of the spatial bins of the galaxies in our sample. The $L_{\rm UV}$ and $L_{\rm IR}$ are derived from the median posteriors of model SEDs obtained from the SED fitting. First, $L_{\rm UV}$ and $L_{\rm IR}$ are calculated for each MCMC sampler chain by integrating in wavelength ranges of $1216-3000$ $\text{\normalfont\AA}$  and $8-1000$ $\mu$m, respectively. Then the median values of $L_{\rm UV}$ and $L_{\rm IR}$ are taken over the whole MCMC sampler chains. The SFR from the total UV and IR luminosities ($\rm{SFR}_{\rm UV+IR}$) is calculated using the prescription from \citet{2005Bell} as follows:
\begin{equation}\label{eq:SFR_UVIR}
\begin{aligned}
& \text{SFR}_{\rm UV+IR}[M_{\odot}\text{yr}^{-1}] \\
& = 1.09 \times 10^{-10} (L_{\rm IR} + 2.2L_{\rm UV}) [L_{\odot}].
\end{aligned}
\end{equation}

\begin{figure*}
\centering
\includegraphics[width=0.48\textwidth]{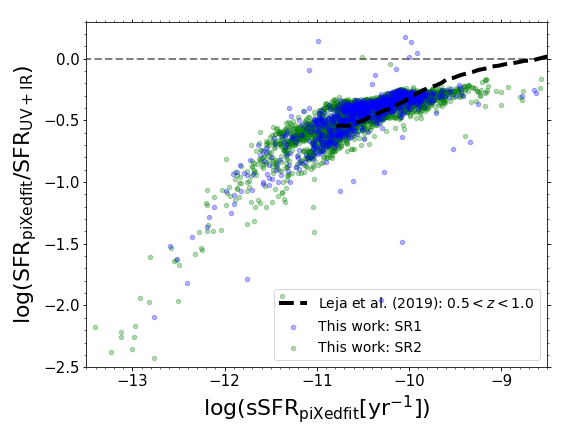}
\includegraphics[width=0.48\textwidth]{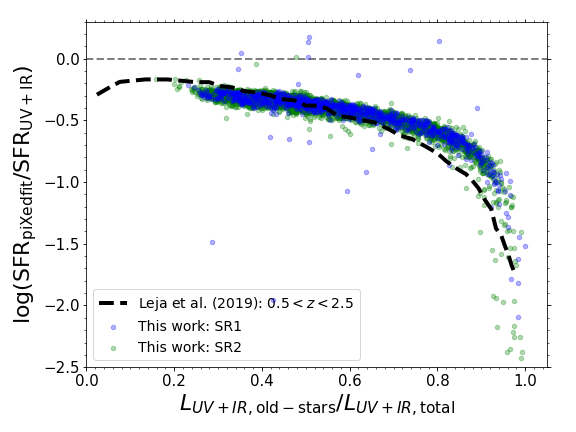}
\caption{Ratios of the SFR derived from \texttt{piXedfit} ($\text{SFR}_{\rm piXedfit}$) and that derived from a prescription that combines UV and IR luminosities ($\text{SFR}_{\rm UV+IR}$) as a function of the sSFR derived from \texttt{piXedfit} ($\text{sSFR}_{\rm piXedfit}$; left panel) and the fractional contribution of stars older than $100$ Myr to the total UV and IR luminosities ($L_{\rm UV+IR, total}$; right panel). $\text{SFR}_{\rm UV+IR}$ is calculated using the prescription from \citet{2005Bell}. The blue and green circles represent the spatial bins from the SR1 and SR2 data cubes, respectively. The black dashed lines represent the trends reported by \citet{2019Leja} who performed SED fitting using \texttt{PROSPECTOR}.}
\label{fig:comp_SFR_piXedfit_UVIR}
\end{figure*}

Figure~\ref{fig:comp_SFR_piXedfit_UVIR}, left panel, shows the ratios of the SFR derived with \verb|piXedfit| ($\text{SFR}_{\rm piXedfit}$) and $\text{SFR}_{\rm UV+IR}$ as a function of the specific SFR ($\text{sSFR}\equiv \text{SFR}/M_{*}$) derived with \verb|piXedfit| ($\text{sSFR}_{\rm piXedfit}$). The spatial bins obtained from the SR1 and SR2 data cubes are shown by the blue and green circles, respectively. It is obvious that $\text{SFR}_{\rm piXedfit}$ is lower than $\text{SFR}_{\rm UV+IR}$ by more than $0.2$ dex for almost all of the spatial bins. There is a trend of increasing ratios (i.e.,~$\text{SFR}_{\rm UV+IR}$ approaching $\text{SFR}_{\rm piXedfit}$) with increasing $\text{sSFR}_{\rm piXedfit}$. In other words, the discrepancy between the two SFR measurements tends to become larger in regions where the star formation activity is weaker (i.e.,~having older stellar populations). A similar trend was also observed by \citet{2019Leja} who fit the integrated UV--MIR SEDs of $0.5<z<2.5$ galaxies using \verb|PROSPECTOR| \citep{2021Johnson}, which also relies on the FSPS code in the modeling of galaxy SED. The black dashed line in the left panel represents the trend reported by \citet{2019Leja} for a sub-sample of galaxies at $0.5<z<1.0$, which is chosen here because this redshift bin is the lowest in their sample. For this trend, which was derived from the median of the overall distribution of the galaxies in the sub-sample of \citet{2019Leja}, $\text{SFR}_{\rm piXedfit}$ and $\text{sSFR}_{\rm piXedfit}$ are replaced with SFR and sSFR derived with the \verb|PROSPECTOR|, respectively. In their sample, star-forming galaxies with $\log(\text{sSFR}[\text{yr}^{-1}])\gtrsim -9.0$ have SFR that agrees with $\text{SFR}_{\rm UV+IR}$. We can see that our result is also consistent with an extension of the trend observed by \citet{2019Leja} toward the less star-forming systems. Our result covers much lower sSFR but does not have coverage in the high sSFR ($\log(\text{sSFR}[\text{yr}^{-1}])\gtrsim -9.3$) because we analyze sub-galactic regions in nearby normal star-forming galaxies which are expected to be less star-forming than the high redshift galaxies observed by \citet{2019Leja}. We emphasize the following three distinctions between our analysis and that of \citet{2019Leja}. (1) Spatial scale: we analyze local kpc-scale region within galaxies, while they focused on properties on a global galaxy-scale; (2) rest-frame wavelength coverage of the photometric data: our data cover FUV--FIR, while their data have a narrower (UV--MIR) coverage; and (3) redshift: it is intriguing to observe that very similar trends are present in both nearby and distant galaxies.     

In order to find the causes of the discrepancy between $\text{SFR}_{\rm piXedfit}$ and $\text{SFR}_{\rm UV+IR}$, we follow \citet{2019Leja} to estimate the contribution of ``old'' ($\text{age}>100$ Myr) stellar populations to the total UV and IR luminosities ($L_{\rm UV+IR} \equiv L_{\rm UV} + L_{\rm IR}$). The age threshold of $100$ Myr is chosen here because it is the age limit adopted by \citet{2005Bell} in modeling the stellar populations to derive the conversion formula (Equation~\ref{eq:SFR_UVIR}). The conversion formula was derived by creating stellar population models with a constant SFR over $100$ Myr. The underlying assumption of the modeling by \citet{2005Bell} is that the young stars that are formed during this star-formation period are the main source of UV (directly) and IR lights (indirectly through the dust heating). 

In our analysis, we calculate the $L_{\rm UV+IR}$ from the stellar populations older than $100$ Myr (hereafter $L_{\rm UV+IR, old-stars}$) based on the posteriors of parameters derived from SED fitting. Briefly speaking, for each MCMC sampler chain in the posteriors, we decompose the model SED into its simple stellar population (SSP) components by following the SFH. To generate the SED of an SSP of a particular age, the stellar mass is defined by the SFH, while the other parameters ($Z_{*}$, dust attenuation parameters, and dust emission parameters) are based on the parameters of the MCMC sampler chain (i.e.,~the composite stellar population, CSP). Then we calculate $L_{\rm UV+IR}$ of each SSP and sum the $L_{\rm UV+IR}$ of SSPs older than $100$ Myr to get the $L_{\rm UV+IR, old-stars}$. After that, we take the median $L_{\rm UV+IR, old-stars}$ over all the MCMC sampler chains. The $L_{\rm UV+IR, total}$, which is the total bolometric $L_{\rm UV+IR}$, is derived in the same way as described above, but the total $L_{\rm UV+IR}$ of each MCMC sampler is taken as a sum over all its SSP components.  

The right panel in Figure~\ref{fig:comp_SFR_piXedfit_UVIR} shows the SFR ratio as a function of the fractional contribution of the old stars to the total $L_{\rm UV+IR}$. The black dashed line shows the result reported by \citet{2019Leja} for their full sample at $0.5<z<2.5$ (as they only performed this analysis for the whole sample). The emerging trend, which agrees with that observed by \citet{2019Leja}, suggests that the SFR discrepancy increases as the fractional contribution of the old stars to the total $L_{\rm UV+IR}$ becomes larger. Overall, the trends in the two panels suggest that the increasing SFR discrepancy in low SFR regions is mainly caused by the increasing contribution of the old stars (which are more abundant in these regions) to the total $L_{\rm UV+IR}$. 

\begin{figure}
\centering
\includegraphics[width=0.48\textwidth]{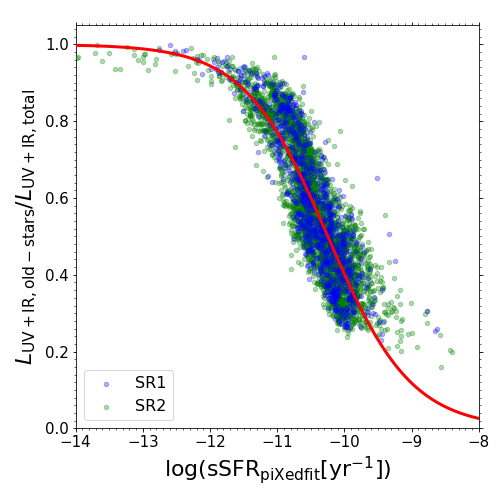}
\caption{The fractional contribution of stars older than 100 Myr to the total UV and IR luminosities as a function of the sSFR derived with \texttt{piXedfit} ($\text{sSFR}_{\rm piXedfit}$). The red line represents the best-fit function in the form of Equation~\ref{eq:fit_LUVIR_oldtot_sSFR} that is obtained using the ODR method.}
\label{fig:ratio_LUVIR_old_tot_vs_sSFR_piXedfit_comb}
\end{figure}

In order to understand how the dust heating by old stars varies with the star formation activity, in Figure~\ref{fig:ratio_LUVIR_old_tot_vs_sSFR_piXedfit_comb}, we show the fractional contribution of the stars older than 100 Myr to the total UV and IR luminosities as a function of $\text{sSFR}_{\rm piXedfit}$. The fractional contributions of the old stars increases with decreasing $\text{sSFR}_{\rm piXedfit}$. To obtain a functional form out of the relation, we fit the distribution of the spatial bins on this plane with the following equation  
\begin{equation} \label{eq:fit_LUVIR_oldtot_sSFR}
y = 0.5\times \text{tanh}(a\log(\text{sSFR}[\text{yr}^{-1}]) + b) + 0.5,
\end{equation} 
where $y$ is $L_{\rm UV+IR,old-stars}/L_{\rm UV+IR, total}$. We use the orthogonal distance regression (ODR) method in the fitting and obtain $a=-0.81\pm 0.01$, $b=-8.26\pm 0.12$ with a scatter of $\sigma=0.08$. This fitting result is shown by the red line in the figure. The uncertainties are estimated using a Monte Carlo method: first we perturb each data point around its original position following a 2D Gaussian distribution with standard deviations dictated by its uncertainties in both axes. Then we perform fitting to the perturbed data. We repeat this process 100 times and estimate the uncertainty by the standard deviations of the obtained $a$ and $b$.   

The exclusion of stars older than $100$ Myr as additional sources of UV and IR light in the modeling of \citet{2005Bell} attributes all the IR luminosity to the heating from young stars. This could overestimate the SFR for given observed UV and IR luminosities. Such an overestimation is expected to be more prominent in less star-forming regions because old stars are more abundant in these regions. Recent studies on the spatially resolved distributions and properties of the stellar populations and dust in nearby galaxies found that a substantial fraction of the energy absorbed by dust originates from the old stellar populations. \citet{2014DeLooze} and \citet{2020Nersesian_b} estimated the fractional contribution of the old stellar populations to the dust heating to be $37$\% and $23$\% in M51, respectively. \citet{2017Viaene} estimated the fraction for M31 to be $91$\%. Analysis of nearby barred galaxies, NGC1365, M83, M95, and M100, by \citet{2020Nersesian_a} found the fractional contributions to the dust heating from the old stellar populations are $\sim32$\%, $36$\%, $53$\%, and $43$\%, respectively. Studies on the integrated SEDs of star-forming galaxies also found a significant contribution ($\sim 40$\%) of old stellar populations on the dust heating \citep[e.g.,][]{2003Hirashita}.

\subsection{Spatially Resolved IRX--$\beta$ Relation} \label{sec:IRX_beta}  
The spatially resolved multiwavelength data sets used in this work provide an opportunity to study the characteristics of dust attenuation and emission at local kpc-scale in nearby spiral galaxies. In principle, the dust attenuation strength can be well estimated using the ratio of the total IR luminosity ($\text{L}_{\rm TIR}$; $3-1100$ $\mu$m) and the FUV luminosity ($\text{L}_{\rm FUV}$; which can be derived from the flux in the FUV band). However, photometry in the infrared bands is not always available. Even when it is available, the spatial resolution is usually low. Another promising way to estimate the dust attenuation strength is to use the so called IRX--$\beta$ relation, which is a relation between the $\text{IRX}\equiv \log(\text{L}_{\rm TIR}/\text{L}_{\rm FUV})$ and the UV spectral slope ($\beta$). Using this relation, the dust attenuation strength can be estimated using $\beta$ alone (which can be constrained relatively easily), without IR data. The above attenuation indicators can then be used for correcting the UV--NIR photometry for the dust attenuation, which is crucial for measuring intrinsic properties of a galaxy. This is one of the reasons why the IRX--$\beta$ relation has been extensively studied observationally as well as theoretically, and motivates us to study it in the spatially resolved scales.   

\begin{figure*}
\centering
\includegraphics[width=0.49\textwidth]{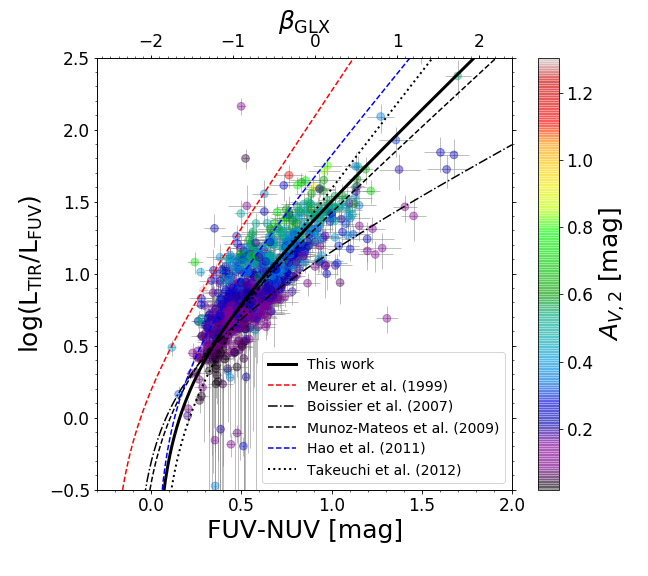}
\includegraphics[width=0.49\textwidth]{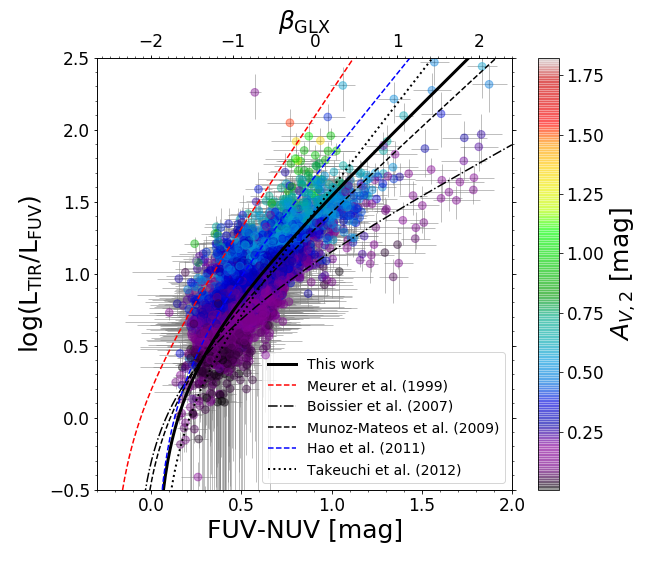}
\caption{The IRX--$\beta$ relation of the spatial bins obtained from the SR1 (left panel) and SR2 (right panel) data cubes. The black solid lines represent the results of fitting a function with the form of Equation~\ref{eq:IRX} to the data points using the ODR method. The color coding are based on $A_{\rm V,2}$. IRX--$\beta$ relations reported by previous studies, which are mostly based on the analysis of integrated SEDs, are also shown for comparison. The corresponding references of those studies are shown in the legend.}
\label{fig:FUVNUV_LTIRFUV_Av2}
\end{figure*}

Previous studies have analyzed the IRX--$\beta$ relation in local galaxies and found a scaling relation that significantly deviates from the original relation proposed by \citet{1999Meurer} for starburst galaxies \citep[e.g.,][]{2007Boissier, 2009Munoz-Mateos_b, 2011Hao, 2012Boquien, 2012Takeuchi}. Taking advantage of our data sets, we study how this relation holds in the spatially resolved scale within our sample galaxies. We plot the IRX--$\beta$ relation of the spatial bins from the SR1 and SR2 data cubes in Figure~\ref{fig:FUVNUV_LTIRFUV_Av2} (left and right panels, respectively). Here we adopt the definition of the UV spectral slope in GALEX bands ($\beta_{\rm GLX}$) given by \citet{2004Kong} as

\begin{equation}
\begin{aligned}
& \beta_{\rm GLX} = \frac{\log(f_{\lambda,\rm FUV})-\log(f_{\lambda,\rm NUV})}{\log(\lambda_{\rm FUV})-\log(\lambda_{\rm NUV})} \\
& = 2.201(\text{FUV}-\text{NUV})-2.
\end{aligned}
\end{equation}  

The observational trends shown in the figure suggest that the $\text{IRX}-\beta$ relation holds even on kpc scales within galaxies. We fit the data with the following equation adapted from \citet[][Equation 13 therein]{2011Hao} and assume an intrinsic $\text{FUV}-\text{NUV}$ color of $0.022$: 
\begin{equation}\label{eq:IRX}
\text{IRX} = \log \left( b \left[ 10^{a [(\text{FUV}-\text{NUV})-0.022]} - 1 \right]  \right),
\end{equation}
where $a$ and $b$ are free parameters to be inferred from the fitting. The intrinsic $\text{FUV}-\text{NUV}$ color of $0.022$ is derived by \citet{2011Hao} from an empirical relationship between the observed $\text{FUV}-\text{NUV}$ color and the attenuation of the $H_{\alpha}$ line for the SINGS sample galaxies. 

We fit the IRX--$\beta$ relations of the SR1 and SR2 data separately using the ODR method. We obtain $a=1.23\pm 0.02$ and $b=2.13\pm 0.06$ for the SR1, while for the SR2 we obtain $a=1.23\pm 0.01$ and $b=2.30\pm 0.05$. As we have done in Section~\ref{sec:comp_SFR_UVIR}, the uncertainties of the derived parameters here are estimated using the Monte Carlo method. These fitting results are shown with the black solid lines in both panels.

The IRX--$\beta$ relations that we obtain from both SR1 and SR2 data sets are consistent (i.e.,~the data points are located on the same locus) with that reported by the previous studies, including \citet{2007Boissier}, \citet{2009Munoz-Mateos_b}, \citet{2011Hao}, and \citet{2012Takeuchi}, as shown by various lines in the figure. All of those studies are based on the analysis of the integrated SEDs of galaxies, except \citet{2009Munoz-Mateos_b}, who analyzed azimuthally-binned SEDs. The majority of the spatial bins in our analysis deviate toward lower IRX and redder $\text{FUV}-\text{NUV}$ colors compared to the IRX--$\beta$ relation of \citet{1999Meurer}, which is originally proposed for starburst galaxies. The color coding is based on the V-band dust attenuation in the diffuse ISM ($A_{\rm V, 2}$) derived from the SED fitting. The trend shows that at a given $\text{FUV}-\text{NUV}$ color, $A_{\rm V, 2}$ tends to increase with increasing IRX. We do not find any other parameters (e.g.,~the dust attenuation power law index, SFR, stellar mass, and age) that have a clear trend along the IRX--$\beta$ relation.

\subsection{Radial Profiles of the Properties of the Stellar Population, Dust, and Gas} \label{sec:radial_profiles}

In this section, we present the resolved distributions, mostly in the form of radial profile, of the stellar, dust, and gas components. Later, in Section~\ref{sec:discussion}, we discuss the relative (i.e.,~ratios) distributions of these components.   

\subsubsection{Radial Profiles of the SFR and Mass Surface Densities of the Stellar, Dust, and Gas Components}\label{sec:rp_SFR_masses}

\begin{figure*}
\centering
\includegraphics[width=0.9\textwidth]{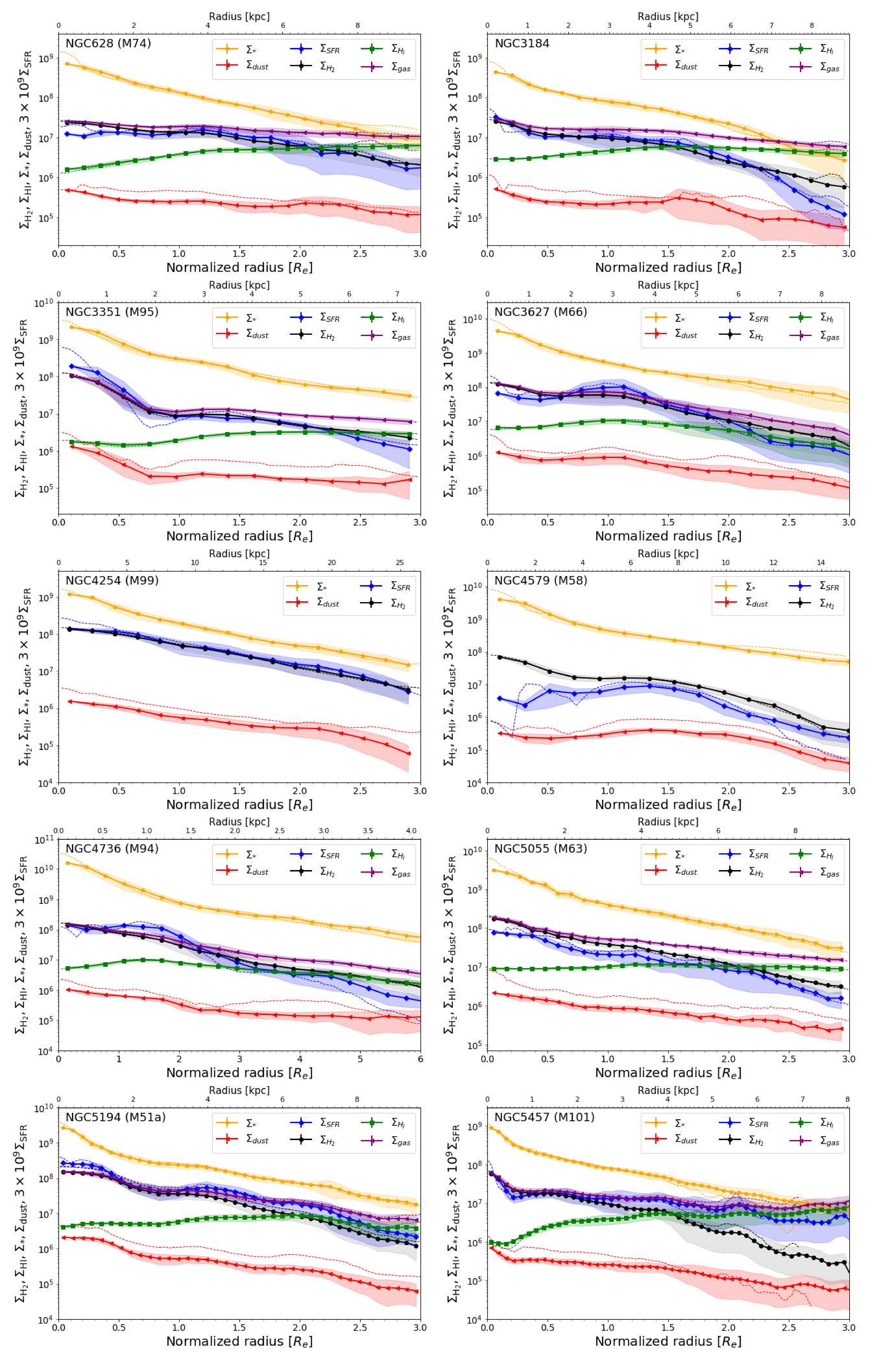}
\caption{Radial profiles of $\Sigma_{\rm SFR}$, $\Sigma_{*}$, $\Sigma_{\rm dust}$, $\Sigma_{\rm HI}$, $\Sigma_{\rm H_{2}}$, and $\Sigma_{\rm gas}$ of the sample galaxies. The $\Sigma_{\rm SFR}(r)$ profiles are multiplied by $3\times 10^{9}$. The $\Sigma_{\rm SFR}(r)$ is in units of $M_{\odot}$ $\rm{yr}^{-1}$ $\rm{kpc}^{-2}$, while the other radial profiles are in units of $M_{\odot}$ $\rm{kpc}^{-2}$.}
\label{fig:collect_radprof_normrad}
\end{figure*}

In order to get a first glance of the radial distributions of the SFR, stellar mass, dust mass, \HI{} mass, $\text{H}_{2}$ gas mass, and total gas mass, we plot the radial profiles of their surface densities ($\Sigma_{\rm SFR}(r)$, $\Sigma_{*}(r)$, $\Sigma_{\rm dust}(r)$, $\Sigma_{\rm HI}(r)$, $\Sigma_{\rm H_{2}}(r)$, and $\Sigma_{\rm gas}(r)$) in one panel. The total gas mass surface density is calculated by $\Sigma_{\rm gas}(r)=\Sigma_{\rm HI}(r)+\Sigma_{\rm H_{2}}(r)$. The radial profiles are derived based on the elliptical apertures with $e$ and PA that are taken from those of the galaxies (see Table~\ref{tab:sample_galaxies}). The radial sampling $\delta r$ is set to be $1$ pixel and the value at each radius is calculated by taking the average of the pixels within the elliptical annulus, while the uncertainty is taken from the standard deviation. We present these radial profiles for all the sample galaxies in Figure~\ref{fig:collect_radprof_normrad}. The $\Sigma_{\rm SFR}(r)$ profiles are multiplied by $3\times 10^{9}$ to bring them into a comparable amplitude with the other radial profiles. The $\Sigma_{\rm SFR}(r)$ is in units of $M_{\odot}$ $\rm{yr}^{-1}$ $\rm{kpc}^{-2}$, while the radial profiles of the other quantities are in $M_{\odot}$ $\rm{kpc}^{-2}$. The radial distance (along the elliptical semi-major axis) is normalized to the half-mass radius (see Table~\ref{tab:sample_galaxies}). The radius is also shown in units of kpc on the top axis of each panel. The radial profiles of all the galaxies, except NGC 4736, extend up to $3R_{e}$. The very small $R_{e}$ of NGC 4736 allows us to study the profiles out to $6R_{e}$. This galaxy has a very compact distribution of $\Sigma_{*}$ with the highest central mass density ($>10^{10}$ $M_{\odot}$ $\rm{kpc}^{-2}$) among the sample galaxies. The radial profiles shown by the thick lines with symbols are based on the SR1 data cubes, while the thinner dashed lines represent the results based on the SR2 cubes. Since we do not have \HI{} maps for NGC 4254 and NGC 4579 (see Section~\ref{sec:sample_select}), these galaxies do not have $\Sigma_{\rm HI}(r)$ and $\Sigma_{\rm gas}(r)$ in our analysis throughout this paper.   

Some general trends revealed from these radial profiles are the following: (1) $\Sigma_{*}(r)$ has the steepest gradient among the radial profiles analyzed here. It has a slightly steeper slope in the central region than in the outskirts, where the profile seems to be exponential. This radial trend reflects the distinct structure between the bulge and disk; (2) the stellar mass is greater than the gas mass over the whole disk in the majority of the galaxies; (3) $\Sigma_{\rm H_{2}}(r)$ increases toward the galactic center (i.e.,~having a negative gradient), while $\Sigma_{\rm HI}(r)$ tends to increase with increasing radius (i.e.,~having a positive gradient). These trends suggest that the molecular gas is centrally concentrated while the \HI{} gas is spatially extended. There is a noticeable transition radius between the \HI{} and $\text{H}_{2}$ gas, which appears to vary from galaxy-to-galaxy; (4) $\Sigma_{\rm SFR}(r)$ broadly follows the overall shape of $\Sigma_{\rm H_{2}}(r)$ in most of the galaxies. This trend is quite noticeable in NGC 3351, NGC 4254, and NGC 5194.

A comparison between the radial profiles derived from the analyses with the SR1 and SR2 data cubes shows no noticeable differences in almost all of the radial profiles, except for $\Sigma_{\rm dust}(r)$, for which the higher spatial resolution data (and less photometric sampling in the FIR; i.e.,~SR2) gives higher dust mass in the entire radial extent. This is likely caused by the exclusion of the two SPIRE bands ($250$ $\mu$m and $350$ $\mu$m) in the SR2 data cubes which can affect the inference of dust mass from the SED fitting. Those two FIR bands are very useful in constraining the shape of the FIR SED, specifically the bump of the dust emission. The unchanged $\Sigma_{\rm SFR}(r)$ and $\Sigma_{*}(r)$ profiles between SR1 and SR2 suggest that these properties are relatively robust against the change in the spatial resolution and exclusion of some photometric data in the FIR.  

Another noticeable trend is found in the barred galaxy NGC 3351 where $\Sigma_{\rm SFR}(r)$, $\Sigma_{\rm H_{2}}(r)$, and $\Sigma_{\rm dust}(r)$ show a break around the radius of $\sim 0.8R_{e}$ and steeply increases toward the galaxy's center. These radial trends agree with the prediction from a zoom-in hydrodynamical simulation of a barred galaxy by \citet{2017Spinoso}. This simulation found that in its early growth phase, the bar causes strong torque, which drives gas infall toward the galaxy center and induces a nuclear starburst. As the bar grows and reaches its maximum length and strength, the gas within its extent ($\sim 2$ kpc) is nearly used up for star formation, causing a break in the gas mass radial profile around the bar extent.      

\subsubsection{Radial Profiles of the Stellar Population Properties}\label{fig:rp_SP}

A stellar population can be characterized by the average age and metallicity ($Z_{*}$) of the member stars. The average age, more precisely the age distribution of the member stars, reflects the star formation history of the system. In this section, we derive the radial profiles of the mass-weighted age and $Z_{*}$ of the sample galaxies. The mass-weighted age, which is derived from the posterior of the star formation history, reflects the average age of stars. Since we adopt a single value of $Z_{*}$ in the SED modeling without modeling the metal enrichment history, the metallicity is directly taken from the fitting result. 

\begin{figure}
\centering
\includegraphics[width=0.5\textwidth]{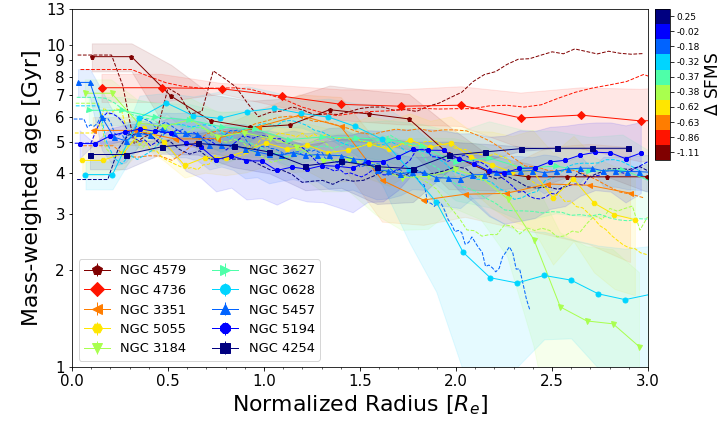}
\includegraphics[width=0.5\textwidth]{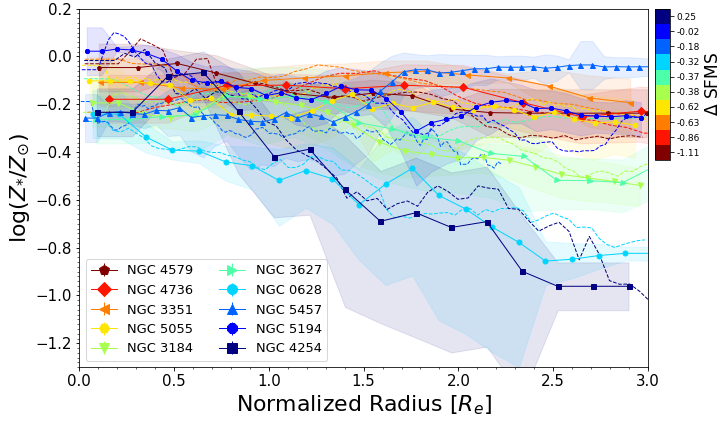}
\caption{Radial profiles of the mass-weighted age (top panel) and stellar metallicity ($Z_{*}$; bottom panel). Different galaxies are represented with different symbols and colors which reflect their distances from the SFMS ridge line ($\Delta$SFMS), as defined in Figure~\ref{fig:plt_sample_SFMS_new}. The solid lines with symbols show radial profiles obtained from the analysis with the SR1 data cubes, while the dashed lines represent the results based on the SR2 data cubes.}
\label{fig:collect_mwage_logzsol}
\end{figure}

Figure~\ref{fig:collect_mwage_logzsol} shows the radial profiles of the mass-weighted age (top panel) and $Z_{*}$ (bottom panel) of the sample galaxies. In contrast to the radial profiles presented in Section~\ref{sec:rp_SFR_masses} which are derived from the values of pixels, these radial profiles are based on the values of spatial bins (i.e.,~pixels within a spatial bin have the same properties). Different galaxies are represented with different symbols and colors that reflect their distances from the SFMS ridge line ($\Delta$SFMS) as defined in Figure~\ref{fig:plt_sample_SFMS_new}. Although NGC 4736 has a small half-mass radius ($R_{e}=0.68$ kpc; see Table~\ref{tab:sample_galaxies}), we present the radial profiles of this galaxy up to $3R_{e}$ in this and the next sections, to be consistent with the other galaxies. Despite the limited coverage to the central $\sim 2$ kpc region of the galaxy, the radial profiles of this galaxy already hint at a systematic radial trend.

Overall, the mass-weighted age and $Z_{*}$ profiles have negative gradients (i.e.,~they decrease with radius) in the majority of the sample galaxies.
The mass-weighted age decreases less rapidly with radius compared to $Z_{*}$. The stellar metallicity shows a more significant variation among the galaxies in the outskirt and intermediate regions than in the central regions. The central regions have nearly solar metallicity ($Z_{\odot}$), while the outskirts have $Z_{*}/Z_{\odot}$ ranging from $\sim 10^{-1} - 1$. The flat trend in the central ($r\lesssim 0.3\times R_{e}$) region in some of the galaxies is caused by the limiting spatial resolution in the bin-wise map of the properties.

The trend with $\Delta$SFMS suggests that the galaxies residing farther below the SFMS ridge line (i.e.,~less star-forming) tend to have stellar populations that are older and more metal rich over the whole disk region. While the radial profiles of mass-weighted age do not show a significant variation in gradient, the radial profiles of $Z_{*}$ of star-forming galaxies tend to have steeper negative gradient than those of passive galaxies. This trend is prominent in NGC 4254, NGC 628, and NGC 5194.

The radial profiles of mass-weighted age and $Z_{*}$ with the SR1 and SR2 data sets show overall similar radial trends for the majority of the sample, which implies that these properties are robust against the variation of spatial resolution and the exclusion of photometric data in the FIR. 

The trends of decreasing stellar ages and metallicities with radius obtained in this analysis are also observed by other recent studies of local galaxies \citep[e.g.,][]{2014Sanchez-Blazquez, 2015GonzalezDelgado, 2017Zheng}. These works used IFS data and performed SED fitting to the spatially resolved spectra. Our derived radial profiles of mass-weighted age and $Z_{*}$ are in good agreement (i.e.,~having similar ranges of value across the entire radial range) with those reported by the above studies. In particular, the mass-weighted ages of spiral galaxies reported by \citet{2015GonzalezDelgado} have ranges of $\sim 3.0-10.0$ Gyr  and $\sim 1.0-8.0$ Gyr in the central and outskirt regions, respectively. The stellar metallicities of spiral galaxies (especially those of Sa to Sbc Hubble types) that they observed have ranges of $\log(Z_{*}/Z_{\odot})\sim -0.1-0.0$ and $\sim -0.6-0.0$ in the central and outskirt regions, respectively. 

The negative gradients of the mass-weighted age and $Z_{*}$ are likely the consequence of the galaxies being assembled in inside-to-outside manner. Models of the hierarchical galaxy formation within the framework of the $\Lambda$CDM cosmology predict a common inside-out scenario of the stellar mass buildup in galaxies \citep[e.g.,][]{1996Kauffmann, 2000Cole, 2002vandenBosch, 2013Aumer}. Because of the inside-out formation, the central region contains more mature stellar populations (i.e.,~older and more metal rich) compared to the outskirt region, which is reflected in the negative gradients of the age and $Z_{*}$ of the stellar populations.

Besides the radial profiles of stellar age and stellar metallicity, there are several other pieces of observational evidence that support the inside-out galaxy formation scenario, in particular, from the studies of the evolution of the radial profiles of $\Sigma_{*}$ in galaxies over a wide range of redshifts \citep[e.g.,][]{2010vanDokkum, 2015Morishita, 2016Nelson, 2016Ibarra-Medel, 2018Abdurrouf}.

\subsubsection{Radial Profiles of the Dust Emission Parameters} \label{sec:rp_dust_emission}

\begin{figure}
\centering
\includegraphics[width=0.5\textwidth]{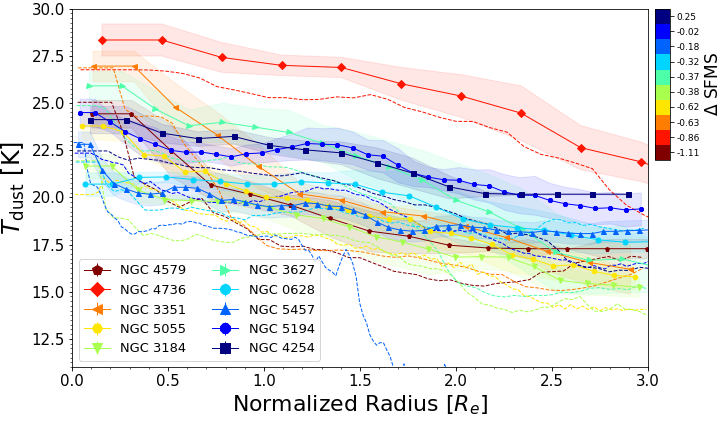}
\includegraphics[width=0.5\textwidth]{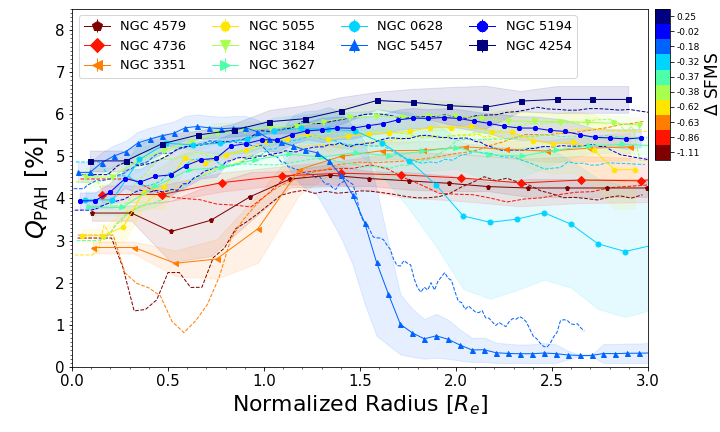}
\includegraphics[width=0.5\textwidth]{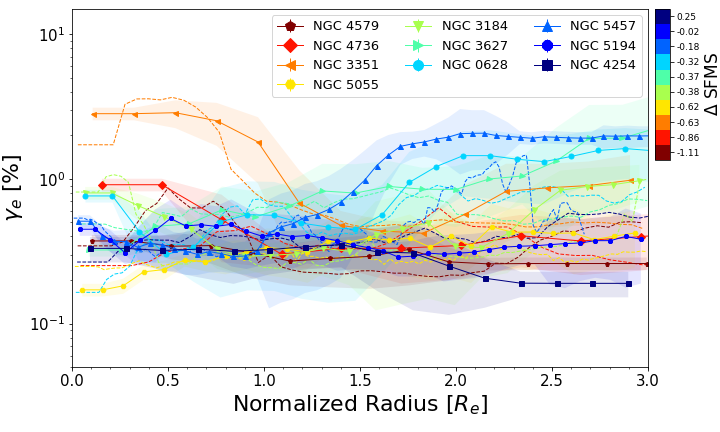}
\caption{Radial profiles of dust emission properties: the dust temperature (top panel), the fraction of PAHs to the total dust mass ($Q_{\rm PAH}$; middle panel), and the relative fraction of dust heated by a minimum starlight intensity $U_{\rm min}$ and that heated by more intense radiation fields with $U_{\rm min}<U\leq U_{\rm max}$ ($\gamma_{e}$; bottom panel). The lines and symbols are the same as those in Figure~\ref{fig:collect_mwage_logzsol}.}
\label{fig:rp_duste_props}
\end{figure}

The dust emission modeling in \verb|piXedfit| is characterized by three parameters: the minimum starlight intensity that heats the dust ($U_{\rm min}$), the PAH abundance ($Q_{\rm PAH}$), and the relative fraction ($\gamma_{e}$) of dust heated by a radiation field with intensity $U_{\rm min}$ (i.e.,~the minimum starlight intensity that heats the dust, which represents the starlight intensity in the diffuse ISM) and that irradiated by more intense radiation fields with $U_{\rm min}<U\leq U_{\rm max}$ (Equation 23 in \citet{2007Draine}). In this section we present the radial profiles of these dust emission parameters in order to gain insight into the dust heating process. We derive the radial profiles of dust temperature (which is calculated using Equations~\ref{eq:Tdust1} and \ref{eq:Tdust2}) instead of $U_{\rm min}$ to directly probe the dust heating state. Figure~\ref{fig:rp_duste_props} shows the radial profiles of dust temperature ($T_{\rm dust}$), $Q_{\rm PAH}$, and $\gamma_{e}$.

The radial profiles of $T_{\rm dust}$ clearly show an increasing trend toward the central region for all the galaxies. The dust temperature in the central region varies from galaxy-to-galaxy with the highest (lowest) being observed in NGC 4736 (NGC 628). The dust temperature across $0<r<3R_{e}$ in NGC 4736 is the highest among the sample galaxies, which may be caused by the AGN in the center. In contrast to the coverage of almost the entire disk region in the other galaxies, the radial profiles of NGC 4736 presented here only extend up to $\sim 2$ kpc because this galaxy has a small half-mass radius. \citet{1999Roberts} observed 12 discrete X-ray sources within the optical region of NGC 4736 with the highest intensity at the nucleus. The second highest dust temperature is found in the central part of NGC 3351, which also shows the most significant increase toward the center. This steep increase of the dust temperature may be caused by the dust heating due to the nuclear starburst activity in this barred galaxy. The central temperature of NGC 5457, which is $\sim 23$ K, is in good agreement with the one obtained by \citet[][Figure 6 therein]{2018Chiang}. There is a noticeable difference between the radial profiles derived with the SR1 and SR2 data cubes in the majority of the galaxies. This is understandable given the importance of the SPIRE $250$ and $350$ $\mu$m bands in constraining the dust temperature. Therefore, the dust temperatures based on the SR1 data sets are expected to be more robust.        

The middle panel in Figure~\ref{fig:rp_duste_props} shows how the PAH abundance varies with radius. Most of the sample galaxies have PAH abundances that are centrally suppressed with broadly flat radial profiles in the disk. In these galaxies, the PAH abundances in the central regions ranges from $\sim 3 \%$ to $5 \%$ of the total dust mass, and those in the disk regions vary from $\sim$4\% to 6\%. This is in good agreement with the $Q_{\rm PAH}$ ($\sim 5 \%$) of nearby spiral galaxies measured by \citet{2007Draine_b}. Exceptions are found in NGC 5457 and NGC 628. The PAH abundance profile in NGC 3351, a barred galaxy, shows a significant decline from $r\sim 1.3\times R_{e}$ to $r\sim 0.6\times R_{e}$, and a slightly increasing trend toward the galactic center. This break in $Q_{\rm PAH}(r)$ happens in the region associated with the bar.

The radial profiles of $\gamma_{e}(r)$ are roughly flat for the majority of the sample galaxies with $\gamma_{e}$ ranges from $\sim 0.2\%$ to $\sim 0.7\%$. Exceptions to this trend are observed in NGC 628, NGC 3351, NGC 3627, and NGC 5457. The radial profile of $\gamma_{e}(r)$ in NGC 3351 is broadly flat in the disk but increases significantly from $r\sim 1.5\times R_{e}$ to $r\sim 1.0\times R_{e}$ and then flattens again toward the galactic center. The other galaxies have radial profiles of $\gamma_{e}(r)$ that roughly decrease from the outskirt region toward the galactic center. The central region of NGC 3351 has the highest $\gamma_{e}$ among the whole regions of the sample galaxies.     

If we look at the color coding, there is no clear correlation between $\Delta$SFMS and the behaviors of $T_{\rm dust}(r)$, $Q_{\rm PAH}(r)$, and $\gamma_{e}(r)$. This suggests that the dust heating process is regulated locally (i.e.,~on kpc scales) and is not strongly related to the global sSFR. A comparison between the SR1 and SR2 data cubes show noticeable discrepancies in $\gamma_{e}(r)$, similar to that observed in $T_{\rm dust}(r)$. Such a difference is expected because these two parameters are mainly constrained by the FIR fluxes. In contrast, consistency is observed between the radial profiles of $Q_{\rm PAH}(r)$ in the two types of data sets. This is not surprising because this parameter is mostly constrained by the MIR fluxes.    

\section{Discussion} \label{sec:discussion} 

Having presented the radial distributions of the absolute properties in the previous section, we are now ready to discuss the relative distributions of the stellar, dust, and gas components. In Section~\ref{sec:rp_sSFR}, we discuss the radial profiles of sSFR, which describes the relative distribution of the SFR and $M_{*}$. We further investigate the radial distributions of the molecular gas mass fraction and depletion time to better understand the spatially resolved sSFR distribution. After that, we investigate the radial profiles of dust-to-stellar mass ratio in Section~\ref{sec:rp_relative_SM_dust} and dust-to-gas mass ratio in Section~\ref{sec:rp_DGR}.   

\subsection{Radial Profiles of sSFR and the Mode of Star Formation} \label{sec:rp_sSFR}

\begin{figure}
\centering
\includegraphics[width=0.5\textwidth]{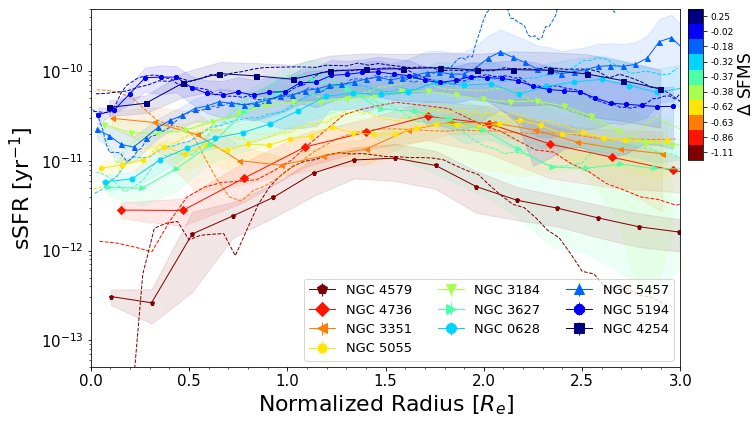}
\caption{Radial profiles of sSFR of the sample galaxies. The lines and symbols are the same as in Figure~\ref{fig:collect_mwage_logzsol}.}
\label{fig:collect_sSFR}
\end{figure}

The global sSFR has been widely used as an indicator of the quenching process in galaxies. A radial profile of sSFR can thus tell us how the quenching process progresses radially across the galaxy. Figure~\ref{fig:collect_sSFR} shows the radial profiles of sSFR. The majority of the sample galaxies have $\text{sSFR}(r)$ that are roughly flat in the disks and suppressed in the central regions. The overall level of sSFR varies significantly from galaxy-to-galaxy, reflecting varying star formation phases among the sample galaxies. There is a clear correlation between the overall amplitude of the $\text{sSFR}(r)$ and the global quantity of $\Delta$SFMS such that more passive galaxies tend to have lower sSFR over the whole disk region and stronger suppression of sSFR at the central region. NGC 3351 has a unique $\text{sSFR}(r)$ which shows a suppression of sSFR around $r \sim R_{e}$ but then turns around to steeply increase toward the center of the galaxy. The elevated SF activity in the center of this galaxy, perhaps in the form of a nuclear starburst, may be caused by the gas infall driven by the torque exerted by the bar \citep[see e.g.,][]{2006Swartz}. There is a noticeable enhancement of sSFR at the intermediate radius ($r\sim R_{e}$) in NGC 3627 which is likely associated with a weak bar structure that connects two spiral arms in this galaxy.

The specific star formation rate can be expressed as
\begin{equation} \label{eq:sSFR}
\text{sSFR} = \frac{\Sigma_{\rm SFR}}{\Sigma_{\rm H_{2}}} \times \frac{\Sigma_{\rm H_{2}}}{\Sigma_{*}} = \frac{1}{t_{\rm dpl,H_{2}}}\times f_{\rm H_{2}},
\end{equation}
where $t_{\rm dpl, H_{2}}$ and $f_{\rm H_{2}}$ are the molecular gas depletion time and the molecular gas-to-stellar mass ratio, respectively. The first factor is called star formation efficieny ($\text{SFE}\equiv 1/t_{\rm dpl, H_{2}}$). Equation~\ref{eq:sSFR} shows that sSFR is controlled by the SFE and the abundance of molecular gas. In order to understand the controlling process of $\text{sSFR}(r)$, specifically what suppresses the sSFR in the central regions of the galaxies and what mechanism likely causes the internal quenching process, we further analyze the radial profiles of $t_{\rm dpl, H_{2}}$ and $f_{\rm H_{2}}$.

\begin{figure}
\centering
\includegraphics[width=0.5\textwidth]{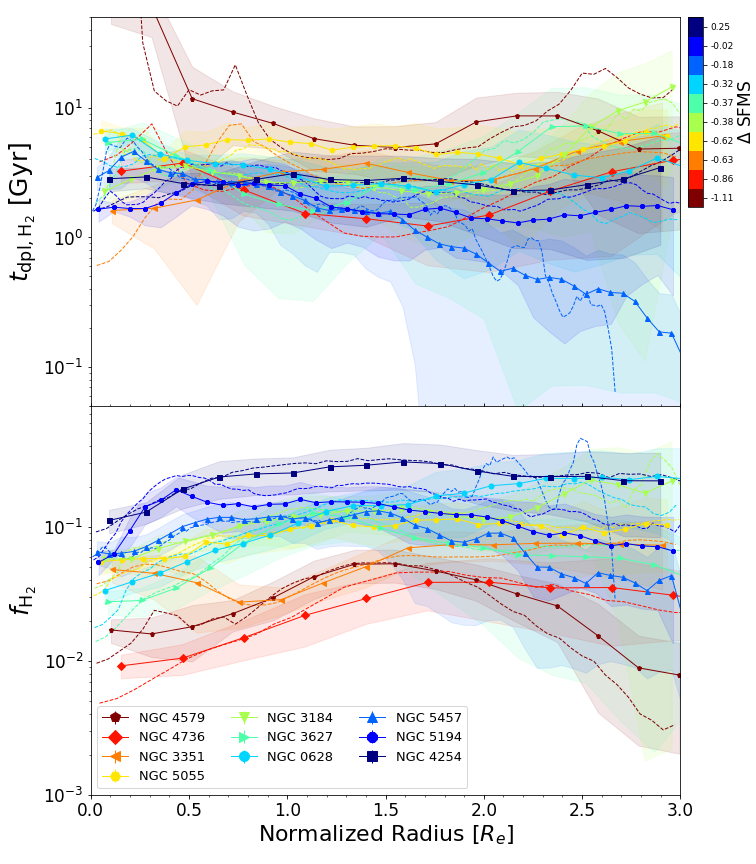}
\caption{Radial profiles of the molecular gas depletion time ($t_{\rm dpl, H_{2}}$; top panel) and the molecular gas-to-stellar mass ratio ($f_{H_{2}}=\Sigma_{\rm H_{2}}/\Sigma_{*}$; bottom panel). The lines and symbols are the same as in Figure~\ref{fig:collect_mwage_logzsol}.}
\label{fig:collect_dplH2_fgas}
\end{figure}

Figure~\ref{fig:collect_dplH2_fgas} shows the radial profiles of $t_{\rm dpl, H_{2}}$ (top panel) and $f_{\rm H_{2}}$ (bottom panel). The radial profiles of $t_{\rm dpl, H_{2}}$ show significantly more variation in the outskirts than in the central regions. The molecular gas depletion time in the outskirt regions ranges from $\sim 0.2$ Gyr to $\sim 15$ Gyr, while in the central regions from $\sim 1$ Gyr to $\sim 10$ Gyr, except for NGC 4579 whose $t_{\rm dpl, H_{2}}$ value is very large, which likely corresponds to the quiescent bulge. The depletion time in the center of NGC 3351 is the shortest among the sample galaxies, indicating a high SFE in the center of this barred galaxy. For the majority of the sample galaxies, $f_{\rm H_{2}}(r)$ shows a suppression in the central regions while having a broadly flat profile in the disks. An exception is found in NGC 3351, where the radial profile shows a suppression at $r\sim 0.8\times R_{e}$ but then turns around toward the center of the galaxy. 

While the profile of $t_{\rm dpl,H_{2}}(r)$ varies from galaxy to galaxy, the shape of $f_{\rm H_{2}}(r)$ seems to be similar among the sample galaxies and looks similar to the shape of the $\text{sSFR}(r)$. Furthermore, we see a clear correlation between the overall level of $f_{\rm H_{2}}(r)$ and $\Delta$SFMS such that less star-forming galaxies tend to have lower $f_{\rm H_{2}}$ over the entire disk region. In contrast, $t_{\rm dpl,H_{2}}(r)$ does not show a clear trend with $\Delta$SFMS. This result may indicate that the local molecular gas content plays a more significant role in regulating the local sSFR compared to the SFE.             

\subsection{What Drives the Suppression of sSFR in the Central Region?} \label{sec:cent_sSFR_driver}

In order to further quantify the relative effects of the SFE and $f_{\rm H_{2}}$ on the suppression of sSFR in the central region, we calculate the ratios of sSFR, $t_{\rm dpl,H_{2}}$, and $f_{\rm H_{2}}$ between the regions inside and outside of the half-mass radius. This is done by measuring the median values of those properties from pixels that lie within the half-mass radius and those lying outside of it. The median statistics are used to avoid the effect of outlier pixels. The uncertainty of the median is estimated using the bootstrap resampling method. 

\begin{figure}
\centering
\includegraphics[width=0.4\textwidth]{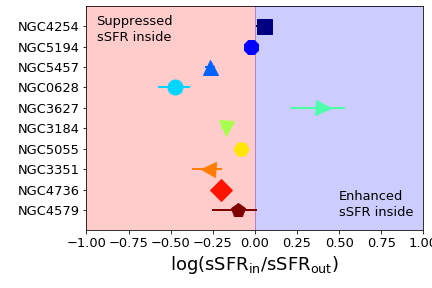}
\includegraphics[width=0.4\textwidth]{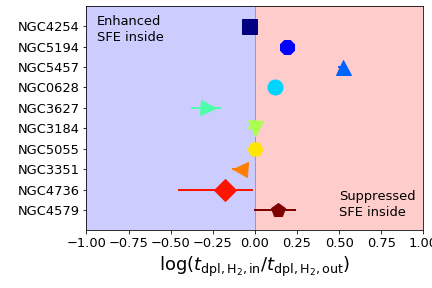}
\includegraphics[width=0.4\textwidth]{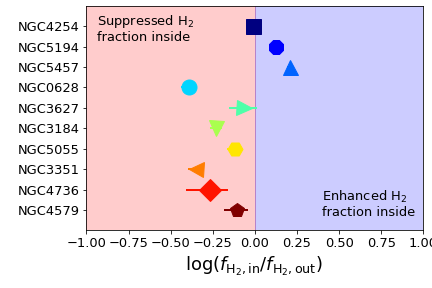}
\caption{Inside-to-outside ratios of the sSFR (top panel), $t_{\rm dpl, H_{2}}$ (middle panel), and $f_{\rm H_{2}}$ (bottom panel). The ratios are taken between the medians (for the pixel values) within and out of the half-mass radius. The red and blue shaded areas indicate a suppression and an enhancement of star formation activity or molecular gas fraction, respectively. The symbols and color-coding of the data points are the same as those in Figure~\ref{fig:plt_sample_SFMS_new}.}
\label{fig:ratio_inout_sSFR_tdplH2_fH2}
\end{figure}

Figure~\ref{fig:ratio_inout_sSFR_tdplH2_fH2} shows the inside-to-outside ratios of the sSFR (top panel), $t_{\rm dpl, H_{2}}$ (middle), and $f_{\rm H_{2}}$ (bottom). The red and blue areas indicate a suppression and an enhancement of star formation activity or molecular gas fraction, respectively. Most of the sample galaxies have suppressed sSFR inside $R_{e}$, as indicated Figure~\ref{fig:collect_sSFR}. The relatively high inside-to-outside ratio of sSFR in NGC 3627 is caused by the enhancement of sSFR around $R_{e}$ (see Figure~\ref{fig:collect_sSFR}). As can be seen from the middle and bottom panels of Figure~\ref{fig:ratio_inout_sSFR_tdplH2_fH2}, the enhanced sSFR in the inner region of this galaxy is more likely caused by an enhancement of SFE in this region rather than by an increase of $f_{\rm H_{2}}$.  

From this analysis, we can see that the causes of the suppression of sSFR in the inner regions of the galaxies are different galaxy-to-galaxy: (1) suppression of the SFE in the inner region, which happens in NGC 5194 and NGC 5457; (2) suppression of $f_{\rm H_{2}}$ in the inner region, which is the case for NGC 3184, NGC 3351, NGC 4736, and NGC 5055; and (3) suppression of both SFE and $f_{\rm H_{2}}$ in the inner region, which occurs in NGC 628 and NGC 4579. These results show that the suppression of $f_{\rm H_{2}}$ is the cause for the suppression of the sSFR in the central regions for the majority of our sample galaxies. In Section~\ref{sec:effect_CO_to_H2_fH2}, we discuss the robustness of this conclusion against the non-constant $\alpha_{\rm CO}$ assumption, especially the $\alpha_{\rm CO}$ that varies with the gas-phase metallicity. Overall, the metallicity-dependent $\alpha_{\rm CO}$ even strengthens this conclusion -- the depletion of molecular gas is the primary cause for the suppression of sSFR at the center.  

The suppression of sSFR in the central regions of galaxies across a wide range of redshift has been observed in many previous studies \citep[e.g.,][]{2014Abramson, 2015Tacchella, 2018Tacchella, 2016GonzalezDelgado, 2017Abdurrouf, 2018Abdurrouf, 2020Abdurrouf, 2018Belfiore, 2018Ellison, 2019Morselli}. However, because of the lack of the high spatial resolution CO data, the physical mechanisms behind the central suppression of sSFR has not been clarified through the analysis of spatially resolved $\text{H}_{2}$ mass and SFE in most of the previous studies. In this work, the high spatial resolution multiwavlength data sets have enabled us to investigate the physical reasons for the suppressed sSFR in the center.

Our result, which proposes that the depletion of molecular gas is the dominant cause for the suppression of sSFR in the central region of galaxies, is in good agreement with the inside-out quenching scenario as demonstrated in the zoom-in cosmological hydrodynamical simulation analyzed by \citet{2016Tacchella}. They found that the galaxies with $\log(M_{*}/M_{\odot})\sim 9.5$ at $2\lesssim z \lesssim 4$ may experience a wet (i.e.,~gas rich) compaction event (due to mergers) that drives a gas transport toward the galaxy's center and triggers a central starburst. This intense star formation in the center, the associated feedback (from e.g.,~an AGN), and the lack of further gas inflow to the center (due to e.g.,~the shock heating of the accreting gas in a dark matter halo that has reached a critical mass of $\sim 10^{12}M_{\odot}$), can cause a depletion of gas in the center and marks the onset of the inside-out quenching process. The central starburst event can build a bulge component in the galaxies. In this work, we observe the galaxies in which the inside-out quenching process has been started a few Gyrs ago and the sSFR as well as the gas mass in their centers already suppressed. In line with this, a recent study using the IllustrisTNG simulation by \citet{2021Nelson} found a significant role of the AGN feedback in the inside-out quenching process in galaxies.

\subsection{Dust-to-stellar Mass Ratio} \label{sec:rp_relative_SM_dust}

\begin{figure}
\centering
\includegraphics[width=0.5\textwidth]{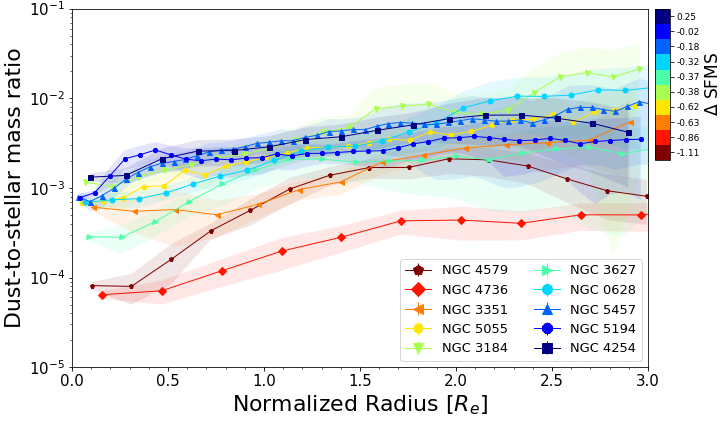}
\caption{Radial profiles of dust-to-stellar mass ratio of the sample galaxies. These radial profiles are obtained from the analysis of the SR1 data cubes.}
\label{fig:collect_rp_dust_to_SM}
\end{figure}

In this section, we study the radial dependence of the dust-to-stellar mass ratio (Figure~\ref{fig:collect_rp_dust_to_SM}). We base this on the SR1 results as their extended wavelength coverage better constrains the dust properties. Overall, the dust-to-stellar mass ratio tends to increase with increasing radius from the galactic center. The dust-to-stellar mass ratios at the galactic centers range from $\sim 6\times 10^{-5}$ to $\sim 10^{-3}$, while those in the outskirt regions range from $\sim 3\times 10^{-4}$ to $\sim 2\times 10^{-2}$. The profile is relatively flat in the disk region of NGC 5194, but it declines toward the galactic center at $r\lesssim 0.5\times R_{e}$. This profile of NGC 5194 is in good agreement with that measured by \citet{2012MentuchCooper}, in the sense that the two radial profiles have a similar range, which is from $\sim 10^{-3}$ to $3\times 10^{-3}$, and show a broadly flat trend in the disk. The radially increasing trend in most of the sample galaxies indicates that $\Sigma_{\rm dust}$ has on average a flatter profile (i.e.,~falling less rapidly with radius) than $\Sigma_{*}$. From the color-coding, we see that there seems to be an indication of lower dust-to-stellar mass ratios in more passive galaxies.  

\subsection{Dust-to-gas Mass Ratio} \label{sec:rp_DGR}

The dust-to-gas mass ratio is an important quantity in studying the physical and chemical processes in the ISM because it tells us about the amount of metals that are locked up in the dust. This quantity also tells us about the relative significance of the dust formation and destruction processes. If dust formation and growth is dominant over dust destruction, the dust-to-gas mass ratio increases with gas-phase metallicity or vice versa \citep[e.g.,][]{1999Hirashita, 2011Hirashita, 2013Asano}. Since the gas-phase metallicity is known to vary with radius, there is expected to be a radial gradient of dust-to-gas ratio.  

\begin{figure}
\centering
\includegraphics[width=0.5\textwidth]{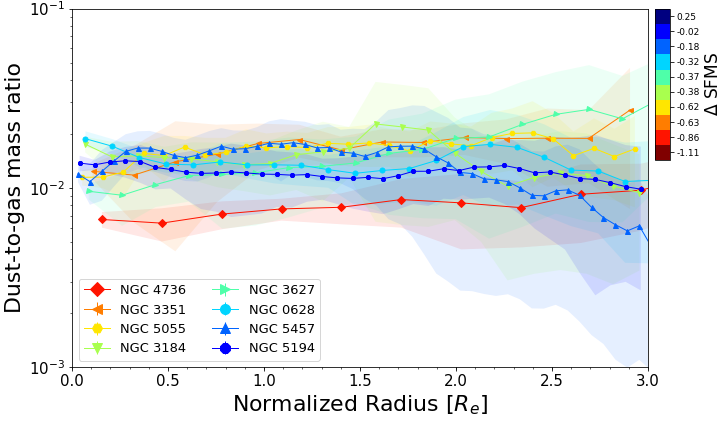}
\caption{Radial profiles of the dust-to-gas mass ratio of the sample galaxies. Only eight galaxies are shown because the other two galaxies (NGC 4254 and NGC 4579) do not have \HI{} data. These radial profiles are obtained from the analysis of the SR1 data cubes.}
\label{fig:collect_rp_DGR}
\end{figure}

Figure~\ref{fig:collect_rp_DGR} shows the radial profiles of dust-to-gas mass ratio. We only analyze eight galaxies that have both \HI{} and $\text{H}_{2}$ data. On average, the radial profiles show broadly constant dust-to-gas mass ratio across the entire radial extent. An exception is observed in NGC 5457 where the dust-to-gas mass ratio increases with decreasing radius from $r\sim 3\times R_{e}$ to $r\sim 1.7\times R_{e}$ and stays roughly flat within $r\sim 1.7\times R_{e}$. The dust-to-gas mass ratios of our sample galaxies range from $\sim 5\times 10^{-3}$ to $\sim 2\times 10^{-2}$. These are roughly similar to the dust-to-gas mass ratio in the Milky Way of $\sim 5\times 10^{-3} - 10^{-2}$, as measured by \citet{1997Sodroski}. The color-coding suggests that there is no clear correlation between the $\Delta$SFMS and the dust-to-gas mass ratio.  

A roughly flat radial profile of dust-to-gas mass ratio is also observed by some previous studies, including \citet{2012Foyle} for NGC 5236 and \citet{2012MentuchCooper} for NGC 5194. The dust-to-gas mass ratio of NGC 5194 that we derive here is consistent with the results of \citet{2012MentuchCooper}, which range from $\sim 10^{-2}$ to $\sim 2\times 10^{-2}$. However, the flat dust-to-gas mass ratio radial profiles are in contradiction with results from other studies. A study of NGC 2403 by \citet{2010Bendo} found a decreasing dust-to-gas mass ratios with radius. \citet{2009Munoz-Mateos_b} also observed that the majority of galaxies in the SINGS sample have a decreasing dust-to-gas mass ratio with radius. Previous studies on the modeling of dust evolution also favor a trend of increasing dust-to-gas mass ratio with increasing gas-phase metallicity or decreasing radial distance from the galactic center \citep[e.g.,][]{2012Mattsson_a, 2020Aoyama}, which suggests dust formation and grain growth dominates over dust destruction in the ISM.

The discrepancy in the radial profiles of the dust-to-gas mass ratio measured by various studies mentioned above is in part caused by the difference in the CO-to-H$_{2}$ conversion factors ($\alpha_{\rm CO}$) adopted by different groups. In addition to the simplest assumption of a constant $\alpha_{\rm CO}$ (as adopted by us and \citealt{2012MentuchCooper}), it is possible to consider a metallicity-dependent $\alpha_{\rm CO}$, which is the case for the studies of \citet{2010Bendo} and \citet{2009Munoz-Mateos_b}. It is thus critical to discuss the effects of varying $\alpha_{\rm CO}$ on the radial profile of dust-to-gas mass ratio to see if the discrepancy is resolved by the treatment of the CO-to-H$_{2}$ conversion factor.

\subsection{Effect of Non-constant CO-to-H$_{2}$ Conversion Factor} \label{sec:effect_CO_to_H2_fH2}

The choice of $\alpha_{\rm CO}$ is one of the caveats (i.e.,~uncertainties) in the study of molecular gas in the ISM. It has been found in previous studies that $\alpha_{\rm CO}$ is not constant, but tends to vary with the ISM environment, especially with the gas-phase metallicity and mass surface density of baryons \citep[e.g.,][]{2013Bolatto, 2013Sandstrom}. Since the ISM environment and the gas-phase metallicity are known to vary with the galactocentric radius, $\alpha_{\rm CO}$ could also vary with radius. Therefore, we expect that this radial variation of $\alpha_{\rm CO}$ would influence some quantities derived in the previous sections that are related to $\Sigma_{\rm H_{2}}$. In this section, we re-calculate the radial profiles of $t_{\rm dpl, H_{2}}$, $f_{\rm H_{2}}$, and dust-to-gas mass ratio using a metallicity-dependent $\alpha_{\rm CO}$. We use this type of $\alpha_{\rm CO}$ instead of the other types (e.g.,~$\alpha_{\rm CO}$ that depends on both gas-phase metallicity and mass surface density of baryons, such as that formulated by \citealt{2013Bolatto}) because of its simplicity.    

We collect radial gradients of the gas-phase metallicity ($12+\log(\text{O}/\text{H})$) of our sample galaxies in the literature. We find radial gradients of $12+\log(\text{O}/\text{H})$ for six galaxies in our sample: four of them (NGC 628, NGC 3184, NGC 5194, and NGC 5457) are taken from \citet{2020Berg} and the other two (NGC 3351 and NGC 5055) are taken from \citet{2014Pilyugin}. In this analysis, we adopt the $\alpha_{\rm CO}$ prescription from \citet{2012Schruba}:
\begin{equation} \label{eq:alpha_CO_metals}
\alpha_{\rm CO}(Z_{\rm gas}) = 8.0\times \left(Z_{\rm gas}/Z_{\odot} \right)^{-2.0},
\end{equation}
where $Z_{\rm gas}$ is the metallicity of the gas, which can be calculated using the following equation adapted from \citet{2018Chiang}
\begin{equation}\label{eq:DTM_2}
Z_{\rm gas} = \frac{\frac{m_{\text{O}}}{m_{\rm H}}10^{(12+\log(\text{O}/\text{H}))-12}}{\frac{M_{\text{O}}}{M_{Z}}\times 1.36},
\end{equation} 
where $m_{\text{O}}$ and $m_{\text{H}}$ are the atomic weights of oxygen and hydrogen, respectively, and $M_{\text{O}}/M_{Z}$ is the mass ratio of oxygen to total metals, which we assume to be $51\%$, following \citet{2003Lodders} who measured it from the solar neighborhood chemical composition.

\begin{figure}
\centering
\includegraphics[width=0.49\textwidth]{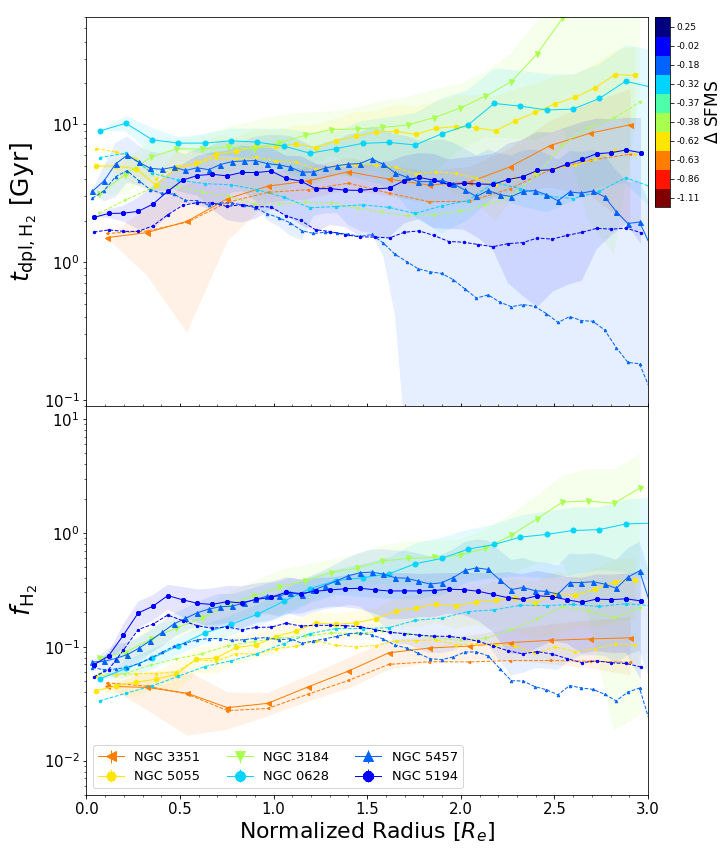}
\caption{The effects of metallicity-dependent $\alpha_{\rm CO}$ conversion factor on the radial profiles of $t_{\rm dpl,H_{2}}(r)$ (top panel) and $f_{\rm H_{2}}(r)$ (bottom panel). In the two panels, the dashed lines with symbols represent radial profiles derived with the constant $\alpha_{\rm CO}$ (the same as in Figure~\ref{fig:collect_dplH2_fgas}), while the solid lines with symbols represent the radial profiles derived using the metallicity-dependent $\alpha_{\rm CO}$ as formulated by \citet{2012Schruba}.}
\label{fig:collect_dpl_fH2_S12}
\end{figure}

\begin{figure}
\centering
\includegraphics[width=0.4\textwidth]{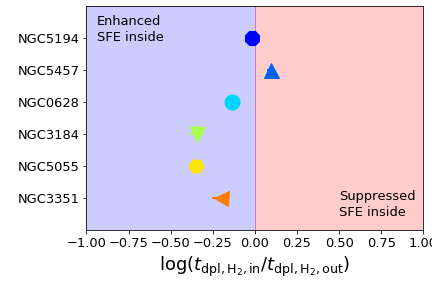}
\includegraphics[width=0.4\textwidth]{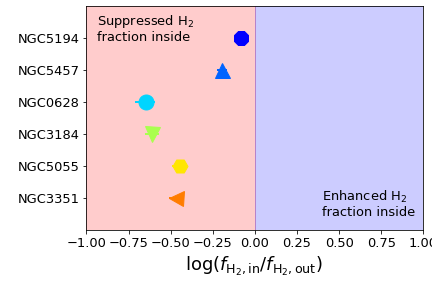}
\caption{Same as Figure~\ref{fig:ratio_inout_sSFR_tdplH2_fH2} for $t_{\rm dpl,H_{2}}$ (top panel) and $f_{\rm H_{2}}$ (bottom panel) but estimated using the metallicity-dependent $\alpha_{\rm CO}$ conversion factor.}
\label{fig:ratio_inout_S12}
\end{figure}

Figure~\ref{fig:collect_dpl_fH2_S12} shows comparisons of the $t_{\rm dpl,H_{2}}(r)$ (top panel) and $f_{\rm H_{2}}(r)$ (bottom panel) between the two cases of $\alpha_{\rm CO}$: the constant $\alpha_{\rm CO}$ (dashed lines with symbols) and the metallicity-dependent $\alpha_{\rm CO}$ (solid lines with symbols and the shaded areas). The metallicity-dependent $\alpha_{\rm CO}$ tends to increase $t_{\rm dpl,H_{2}}(r)$ and $f_{\rm H_{2}}(r)$ at all radii by a factor that gradually increases with radius. This makes these radial profiles flatter or even rising with increasing radius (i.e.,~having positive gradient), which is prominent in the new $f_{\rm H_{2}}(r)$. Figure~\ref{fig:ratio_inout_S12} shows the inside-to-outside ratios of the $t_{\rm dpl, H_{2}}$ (top panel) and the $f_{\rm H_{2}}$ (bottom panel) derived using the metallicity-dependent $\alpha_{\rm CO}$. As we can see, all six galaxies now have centrally-suppressed $f_{\rm H_{2}}(r)$. NGC 5194 and NGC 5457 have centrally-enhanced $f_{\rm H_{2}}(r)$ when a constant $\alpha_{\rm CO}$ is assumed (see Section~\ref{sec:rp_sSFR} and Figure~\ref{fig:ratio_inout_sSFR_tdplH2_fH2}). With the metallicity-dependent $\alpha_{\rm CO}$, the suppression of $f_{\rm H_{2}}$ is now the major cause of the suppression of sSFR in the central region. This scenario applys to all but one galaxy: NGC 5457 where the both scenarios seem to work: the suppression of $f_{\rm H_{2}}$ and the SFE.              

\begin{figure}
\centering
\includegraphics[width=0.49\textwidth]{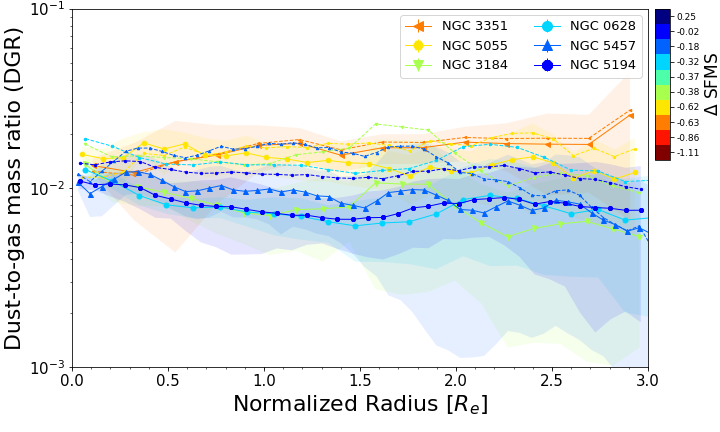}
\caption{Comparison between the radial profiles of dust-to-gas mass ratio derived with an assumption of the constant $\alpha_{\rm CO}$ and that derived with the metallicity-dependent $\alpha_{\rm CO}$ prescription of \citet{2012Schruba}. The lines and symbols are the same as in Figure~\ref{fig:collect_dpl_fH2_S12}.}
\label{fig:collect_DGR_aCO_MW_vs_S12_P14}
\end{figure}

Figure~\ref{fig:collect_DGR_aCO_MW_vs_S12_P14} shows a comparison of the radial profiles of the dust-to-gas mass ratio derived with the two assumptions of $\alpha_{\rm CO}$. The metallicity-dependent $\alpha_{\rm CO}$ tends to reduce the dust-to-gas mass ratio by a factor that increases with radius, making it gradually decrease with radius (i.e.,~having a negative gradient). These new radial profiles agree with the majority of previous observational \citep[e.g.,][]{2009Munoz-Mateos_b, 2010Bendo} and theoretical \citep[e.g.,][]{2012Mattsson, 2012Mattsson_a, 2020Aoyama} studies. The decreasing dust-to-gas mass ratio with radius, which also implies an increasing dust-to-gas mass ratio with the gas-phase metallicity (since the gas-phase metallicity has a negative gradient), indicates dust formation and grain growth dominates over dust destruction in the ISM.                

The above results show a significant influence of $\alpha_{\rm CO}$ on the radial profiles of the properties that are related to the molecular gas. In a forthcoming paper, we will apply various $\alpha_{\rm CO}$ prescriptions in the literature to discuss how $\alpha_{\rm CO}$ influences the radial profiles of the dust-to-gas mass ratio as well as those of the dust-to-metals ratio.    

\section{Summary and Conclusions} \label{sec:summary}

We investigate the spatially resolved distributions and properties of the stars, dust, and gas in ten nearby spiral galaxies. The sample galaxies are in the star-forming and green-valley phases (Table~\ref{tab:sample_galaxies}). We make use of multiwavelength imaging data of 20-24 bands taken from the GALEX, SDSS, 2MASS, WISE, \textit{Spitzer Space Telescope}, and \textit{Herschel Space Observatory}. To obtain the spatially resolved properties of the stellar population and dust, we conduct spatially resolved FUV--FIR SED fitting using our newly-developed software \verb|piXedfit|, while for deriving the spatially resolved surface mass densities of the neutral and molecular gas ($\Sigma_{\rm HI}$ and $\Sigma_{\rm H_{2}}$, respectively) we use 21cm \HI{} and CO data from the THINGS and HERACLES surveys.

The key results are the following:
\begin{enumerate}
\item The spatially resolved SFRs derived from \verb|piXedfit| are significantly lower than those obtained from a prescription that combines the UV and IR luminosities ($\text{SFR}_{\rm UV+IR}$). This discrepancy can be explained by a significant contribution from the stellar populations older than $100$ Myr to the dust heating, which is not corrected in the prescription of $\text{SFR}_{\rm UV+IR}$. This effect is more significant in the passive systems (both on galactic and kpc-scales) where old stars are more abundant.

\item We use two different sets of imaging data (FUV--SPIRE $350$ $\mu$m and FUV--PACS $160$ $\mu$m) to investigate the effects of varying wavelength coverage and spatial resolution on the derived spatially resolved properties of the stellar population and dust. We find that some properties related to dust (dust mass, dust temperature, $A_{V,1}$, and $\gamma_{e}$) are sensitive to the exclusion of the SPIRE bands, while other properties ($M_{*}$, SFR, mass-weighted age, $Z_{*}$, $A_{V,2}$, and $Q_{\rm PAH}$) are not affected by this exclusion.

\item The spatially resolved IRX--$\beta$ relation is consistent with that observed on the global scale by previous studies. The resolved IRX--$\beta$ relation deviates from the original integrated IRX--$\beta$ relation reported by \citet{1999Meurer} for starburst galaxies.

\item The radial surface density profiles of SFR and molecular gas are roughly similar in shape. The radial surface density profiles of dust and total gas (i.e.,~total of atomic and molecular gas) also show a similarity in shape. The radial profiles of $\Sigma_{\rm H_{2}}$ are centrally concentrated, while those of $\Sigma_{\rm HI}$ are more spatially extended. 

\item The stellar mass-weighted ages and metallicities ($Z_{*}$) both decrease with increasing radial distance from the galactic center (i.e.,~have a negative gradient). Less star-forming galaxies (i.e.,~those that reside farther below the SFMS ridge line on the SFR--$M_{*}$ diagram) tend to have stellar populations that are older and more metal rich over the whole disk. Moreover, the star-forming galaxies tend to have steeper $Z_{*}$ gradient than passive galaxies. These trends are consistent with the inside-out galaxy formation scenario.  

\item The dust temperature decreases with increasing radius in all of the sample galaxies, while the PAH abundance is roughly constant in the disk and suppressed in the central region in the majority of them. The fraction of the dust mass that is exposed to a more intense local radiation field than the mean diffuse radiation, $\gamma_{e}$, appears to be constant across the entire radius for most of the sample galaxies. An exception is found in the barred galaxy NGC 3351 where $\gamma_{e}(r)$ is centrally-peaked and has a high value ($\sim 3\%$) in the center. 

\item The sSFR radial profile of most of the sample galaxies is suppressed in the central regions while roughly constant in the disks. This is consistent with the inside-out quenching scenario. To find the causes of the sSFR suppression in the central regions, we investigate the radial profiles of the molecular gas depletion time ($t_{\rm dpl, H_{2}}$) and the molecular gas-to-stellar mass ratio ($f_{\rm H_{2}} \equiv \Sigma_{\rm H_{2}}/\Sigma_{*}$). The molecular gas depletion time varies from galaxy-to-galaxy, ranging from $\sim 2$ to $11$ Gyr. The molecular gas-to-stellar mass ratio in the majority of the sample galaxies shows a suppression in the central region and is found to be roughly constant in the disk. The causes of the suppression of sSFR in the central region vary from galaxy to galaxy: (1) suppression of the star formation efficiency (SFE) in the central region (NGC 5194 and NGC 5457); (2) suppression of the $f_{\rm H_{2}}$ in the central region (NGC 3184, NGC 3351, NGC 4736, and NGC 5055); and (3) suppression of both the SFE and $f_{\rm H_{2}}$ (NGC 628 and NGC 4579). Applying a metallicity-dependent CO-to-H$_{2}$ conversion factor ($\alpha_{\rm CO}$) instead of the constant one makes the suppression of $f_{\rm H_{2}}$ in the center even more stronger: in this case, lack of $\text{H}_{2}$ is the primary driver for the inside-out star formation quenching process.

\item The dust-to-stellar mass ratio tends to increase with increasing radius, while the dust-to-gas mass ratio is broadly constant across the entire radial extent. However, the profile of dust-to-gas mass ratio is significantly influenced by the assumption of $\alpha_{\rm CO}$. A metallicity-dependent $\alpha_{\rm CO}$ tends to make the dust-to-gas mass ratio to decrease with radius (i.e.,~have a negative gradient), which agrees with the majority of previous studies (both observation and simulation). The increasing dust-to-gas mass ratio with decreasing radius (i.e.,~increasing gas-phase metallicity) suggests dust formation and grain growth dominates over dust destruction in the ISM.      
    
\end{enumerate}

With this work, we also demonstrate the capabilities of \verb|piXedfit| in various aspects related to the analysis of the spatially resolved SEDs of galaxies, including the image processing, pixel binning, and SED fitting. As demonstrated in Appendix~\ref{sec:SEDfit_test_mock}, the SED fitting methods adopted in \verb|piXefit| can provide robust estimates of the properties of stellar population and dust. \verb|piXedfit| is available at \url{https://github.com/aabdurrouf/piXedfit}.

\begin{acknowledgments}
We thank the anonymous reviewer who has given valuable comments that helped to improve this paper. We thank I-Da Chiang for valuable discussion and comments. We are grateful for support from the Ministry of Science and Technology of Taiwan under grants MOST 108-2112-M-001-011 and MOST 109-2112-M-001-005 and a Career Development Award from Academia Sinica (AS-CDA-106-M01). HH thanks the Ministry of Science and Technology of Taiwan for support through grant MOST 107-2923-M-001-003-MY3 and MOST 108-2112-M-001-007-MY3, and the Academia Sinica for Investigator Award AS-IA-109-M02. S.T. is supported by the 2021 Research Fund 1.210134.01 of UNIST (Ulsan National Institute of Science \& Technology). PFW acknowledges the support of the fellowship from the East Asian Core Observatories Association. The computations in this research were run on the TIARA cluster at ASIAA. 

This research made use of Astropy\footnote{\url{http://www.astropy.org}}, a community-developed core Python package for Astronomy \citep{2013Astropy, 2018Astropy}. This research made use of \texttt{Photutils}, an \texttt{Astropy} package for detection and photometry of astronomical sources \citep{2019Bradley}. 

This work is based on observations made with the NASA Galaxy Evolution Explorer (GALEX), which is operated for NASA by the California Institute of Technology under NASA contract NAS5-98034. 

Funding for the Sloan Digital Sky Survey IV has been provided by the Alfred P. Sloan Foundation, the U.S. Department of Energy Office of Science, and the Participating Institutions. SDSS-IV acknowledges support and resources from the Center for High-Performance Computing at the University of Utah. The SDSS web site is www.sdss.org. SDSS-IV is managed by the Astrophysical Research Consortium for the Participating Institutions of the SDSS Collaboration including the Brazilian Participation Group, the Carnegie Institution for Science, Carnegie Mellon University, the Chilean Participation Group, the French Participation Group, Harvard-Smithsonian Center for Astrophysics, Instituto de Astrof\'isica de Canarias, The Johns Hopkins University, Kavli Institute for the Physics and Mathematics of the Universe (IPMU)/University of Tokyo, the Korean Participation Group, Lawrence Berkeley National Laboratory,  Leibniz Institut f\"ur Astrophysik Potsdam (AIP), Max-Planck-Institut f\"ur Astronomie (MPIA Heidelberg), Max-Planck-Institut f\"ur Astrophysik (MPA Garching), Max-Planck-Institut f\"ur Extraterrestrische Physik (MPE), National Astronomical Observatories of China, New Mexico State University, New York University, University of Notre Dame, Observat\'ario Nacional / MCTI, The Ohio State University, Pennsylvania State University, Shanghai Astronomical Observatory, United Kingdom Participation Group, Universidad Nacional Aut\'onoma de M\'exico, University of Arizona, University of Colorado Boulder, University of Oxford, University of Portsmouth, University of Utah, University of Virginia, University of Washington, University of Wisconsin, Vanderbilt University, and Yale University. 

This publication makes use of data products from the Two Micron All Sky Survey, which is a joint project of the University of Massachusetts and the Infrared Processing and Analysis Center/California Institute of Technology, funded by the National Aeronautics and Space Administration and the National Science Foundation. 

This publication makes use of data products from the Wide-field Infrared Survey Explorer, which is a joint project of the University of California, Los Angeles, and the Jet Propulsion Laboratory/California Institute of Technology, funded by the National Aeronautics and Space Administration. 

This work is based [in part] on observations made with the Spitzer Space Telescope, which was operated by the Jet Propulsion Laboratory, California Institute of Technology under a contract with NASA. 

Herschel is an ESA space observatory with science instruments provided by European-led Principal Investigator consortia and with important participation from NASA.
\end{acknowledgments}

%

\vspace{5mm}
\facilities{GALEX, Sloan, FLWO:2MASS, WISE, \textit{Spitzer}, \textit{Herschel}, IRAM:30m, VLA}


\software{
\texttt{piXedfit} \citep{2021Abdurrouf2, 2021Abdurrouf}
\texttt{Astropy}  \citep{2013Astropy, 2018Astropy},
\texttt{Photutils}  \citep{2019Bradley},
\texttt{reproject} \citep{2018robitaille},
\texttt{SExtractor}  \citep{1996bertin},
\texttt{Montage} \citep{2010Jacob},
\texttt{FSPS}  \citep{2009Conroy,2010Conroy},
\texttt{emcee}  \citep{2013Foreman},
\texttt{matplotlib} \citep{2007Hunter},
\texttt{NumPy} \citep{2020Harris},
}



\appendix

\section{Testing the SED fitting performances of \texttt{piXedfit} with mock SEDs} \label{sec:SEDfit_test_mock}

In order to test the performances of \verb|piXedfit| in terms of parameter inferences, we conduct SED fitting tests using semi-empirical mock SEDs. First, we generate 200 semi-empirical mock SEDs using FSPS with parameter values taken from the measured values of the spatial bins of the galaxies in our sample. Specifically, we select the first 20 bins of each galaxy (but excluding the central bin to avoid AGN contamination), and use the measured values of the parameters (from the SED fitting analysis that is performed for the main analysis presented in this paper) as references in generating the mock SEDs. We choose to use the actual measured values for these parameters, rather than randomly drawing values from a pre-defined range in order to create more physically-motivated mock SEDs. In the case of randomly drawn parameters, we can not be sure that the combinations of those parameters exist in nature. Following our fitting setup in the analysis of this paper (see Section~\ref{sec:sed_fitting}), we assume the \citet{2003Chabrier} IMF, a double power law SFH, and the \citet{2000Charlot} dust attenuation law in generating the mock SEDs. In these tests, we turn off the AGN component because we aim at testing the inference of parameters associated with the stellar population and dust. We convolve the mock spectra with the same set of 24 photometric filters as used in the main analysis of this paper (see Table~\ref{tab:filters_set}). We mimic the observational noises by injecting artificial noises to the mock SEDs assuming a Gaussian distribution with $\rm{S}/\rm{N}$ of $10$ in all bands. 

\begin{figure*}
\centering
\includegraphics[width=0.95\textwidth]{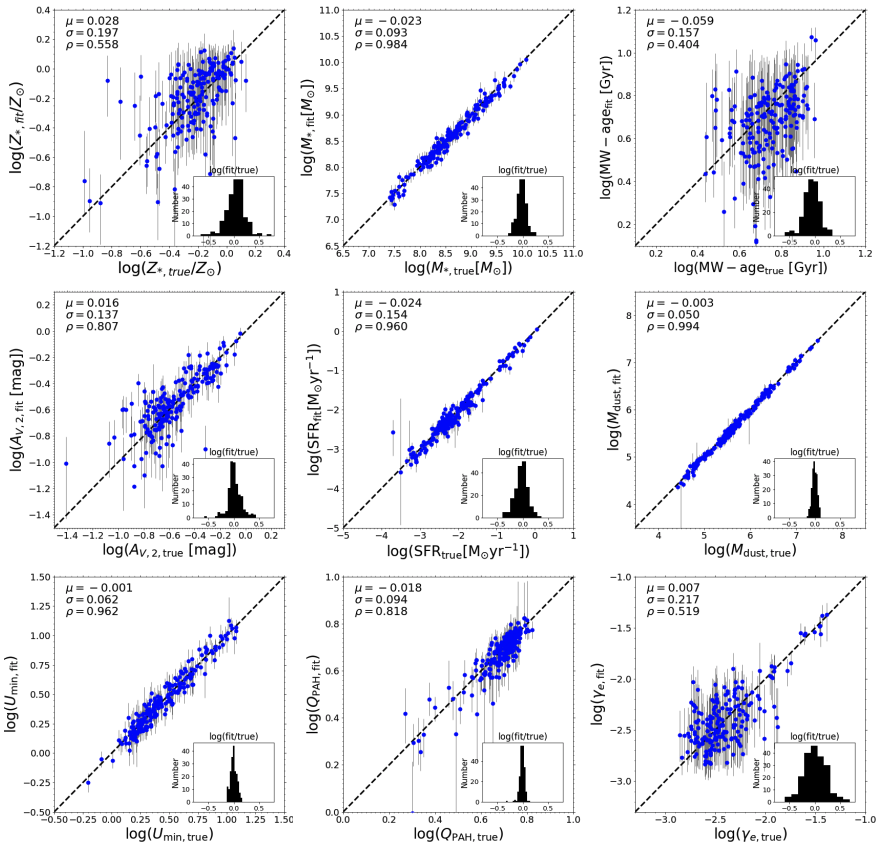}
\caption{Comparisons between the values of parameters derived from SED fitting and the true values of the mock SEDs. Panels from left to right in the first row to the third row show metallicity ($Z_{*}$), $M_{*}$, mass-weighted age, dust attenuation in the diffuse ISM ($A_{V,2}$), SFR, dust mass ($M_{\rm dust}$), $U_{\rm min}$, $Q_{\rm PAH}$, and $\gamma_{e}$.}
\label{fig:fittests}
\end{figure*}

We fit the mock SEDs using the MCMC method, assuming the same setup (IMF, SFH form, and dust attenuation law) as that used in generating the mock SEDs. Figure~\ref{fig:fittests} shows comparisons between the fitting results and the true parameters of mock SEDs for 9 parameters: $Z_{*}$, $M_{*}$, mass-weighted age, $A_{V,2}$, SFR, $M_{\rm dust}$, $U_{\rm min}$, $Q_{\rm PAH}$, and $\gamma_{e}$. The comparisons show that the true parameters are well recovered, indicated by the low absolute values of the offset ($\mu<0.06$ dex), scatter ($\sigma<0.22$ dex), and the high value of the Spearman's rank-order correlation coefficient ($\rho>0.4$). The best-recovered parameter is $M_{\rm dust}$, followed by $U_{\rm min}$, $M_{*}$, and SFR. Although we only use the photometric data, interestingly, $Z_{*}$ is well recovered by the fitting results. Previous studies found that it is difficult to recover $Z_{*}$ and mass-weighted age using SED fitting with UV--NIR photometry \citep[see e.g.,~][and reference therein]{2021Abdurrouf, 2021Tacchella}. This success is in part supported by the wide wavelength coverage of the photometry used in the current analysis, which could break the degeneracy among $Z_{*}$, dust attenuation, and age. The comparisons also show that the dust attenuation parameters ($U_{\rm min}$, $Q_{\rm PAH}$, $\gamma_{e}$, and $M_{\rm dust}$) are recovered very well, suggesting the robustness of the dust properties derived with \verb|piXedfit|.      

\bibliography{nearbygals1}{}
\bibliographystyle{aasjournal}



\end{document}